\documentclass[twocolappendix,twocolumn]{aastex7}

\newcommand{\Msun}{\mbox{\,$\rm M_{\odot}$}}        
\newcommand{\Lsun}{\mbox{\,$\rm L_{\odot}$}}        

\usepackage{tabularx}
\usepackage{amsmath,amssymb}
\usepackage{lineno}
\usepackage{ulem}
\graphicspath{{./}{PDF/}}
\linenumbers

\begin{document}

\title{Theoretical models for the Late Thermal Pulse in post-AGB stars: the case of DY Cen}

\author[0009-0005-2920-7261]{Zhongyang Liu}
\affiliation{School of Physics and Astronomy,
Beijing Normal University,
Beijing, 100875, China}
\email{jerry@mail.bnu.edu.cn}

\author[0000-0003-1759-0302]{C. Simon Jeffery}
\affiliation{Armagh Observatory and Planetarium, College Hill, Armagh BT61 9DG, UK 
}
\email{simon.jeffery@armagh.ac.uk}

\author[0000-0002-3672-2166]{Xianfei Zhang}
\affiliation{School of Physics and Astronomy,
Beijing Normal University,
Beijing, 100875, China}
\email{zxf@bnu.edu.cn}

\author[0000-0001-7566-9436]{Shaolan Bi}
\affiliation{School of Physics and Astronomy,
Beijing Normal University,
Beijing, 100875, China}
\email{bisl@bnu.edu.cn}

\author[0000-0001-6396-2563]{Tanda Li}
\affiliation{School of Physics and Astronomy,
Beijing Normal University,
Beijing, 100875, China}
\email{litanda@bnu.edu.cn}

\correspondingauthor{Xianfei Zhang}
\email{zxf@bnu.edu.cn}

\correspondingauthor{C. Simon Jeffery}
\email{simon.jeffery@armagh.ac.uk}

\begin{abstract}
We present theoretical predictions of the born-again scenario for post-asymptotic giant-branch stars. 
An extensive model grid for born-again objects has been constructed, particularly including models for the Very Late Thermal Pulse with and without convective overshooting, and also including models for the Late Thermal Pulse.  
We constructed a large parameter space to analyze the dependencies of the born-again model on core mass, hydrogen-envelope mass, and overshoot parameters, and we analyzed how changes in these parameters affect the models' evolution.
We applied our grid of models to interpret observations of DY\,Cen, a star exhibiting characteristics similar to confirmed born-again stars.
We compared DY\,Cen with models from multiple aspects, including heating rate, evolutionary tracks, and surface abundances.
Ultimately, we concluded that none of our born-again models could match all of the observed properties of DY\,Cen, especially its surface chemistry; DY\,Cen is therefore an unlikely born-again star.
\end{abstract}


\section {Introduction} \label {s: intro}

\subsection{The Asymptotic Giant Branch and beyond}

Asymptotic giant branch (AGB) stars are low- and intermediate-mass stars that have completed core helium burning and so consist of a degenerate carbon-oxygen core, a helium-rich inter-shell, and a hydrogen-rich envelope. 
For the most part, their high luminosity is maintained by nuclear hydrogen burning at the base of the hydrogen envelope.
This is interrupted by unstable helium burning at the base of the helium inter-shell in a process known as a ``thermal pulse''. 

Once a star has lost the most of its envelope, it departs the AGB and enters the post-asymptotic giant branch (post-AGB). 
In this post-AGB phase, stars contract, maintaining roughly the same luminosity, until they reach a planetary nebula or hot pre-white dwarf phase, after which they start to dim and cool as white dwarfs.

Before leaving the AGB, stars exhibit very strong mass loss associated with their high luminosities.  
Although still dominated by hydrogen, the surface abundances of helium, carbon, oxygen, nitrogen, and other elements can be significantly enhanced by ``third dredge-up'', a process by which convection brings fresh nuclear products produced in a thermal pulse to the surface. 

Leaving the AGB does not mean that the hydrogen shell switches off completely, \textbf{when the hydrogen envelope drops below a critical mass \citep{paczynski71,miller_2016,Kwok2024}the envelope must contract.}
Since the hydrogen-shell can still add helium to the inter shell, and contraction will compress and heat the inter-shell, the latter may therefore ignite helium if and when the inter-shell reaches a critical temperature and density, even though the star has become a dwarf or white dwarf \citep{schoenberner79,iben84}.
Such ignition therefore depends on the state of the inter-shell at the point where the hydrogen-shell could no longer sustain the star on the AGB, as well as on the residual hydrogen-mass which controls the rate at which hydrogen is being added to the inter-shell. 
If it occurs, the ignition, burning and expansion cycle is known  as a ``last thermal pulse'', or ``born-again phenomenon'' \citep{Iben_1984b}, and can substantially reduce the hydrogen content at the stellar surface. 
This paper is concerned with modeling that process, in particular with respect to understanding the hydrogen-deficient star DY\,Cen. 

\subsection{The Late Thermal Pulse in the evolution of post-AGB stars}

\cite{Blocker_2001} identified three types of ``last thermal pulse'':  
an AGB Final Thermal Pulse (AFTP) occurring at the very end of the AGB evolution, a Late Thermal Pulse (LTP) occurring during the post-AGB evolution when hydrogen burning is still on, and a Very Late Thermal Pulse (VLTP) occurring on the WD when hydrogen burning has already ceased. 
In VLTP and LTP models, the star is forced to expand to become a cool supergiant where it may stay for some time; it then contracts along a post-AGB track for a second time and finally reaches the white dwarf cooling track.
This evolution may be fast, taking months to tens of years to expand and roughly ten times that to contract. 
This is also an important observational characteristic for verifying that the star is a ``born-again star''.

A classical method to distinguish whether a star experienced a LTP or a VLTP was to check if the surface retains its hydrogen-rich primordial composition. 
If the surface shows no significant change, the star likely experienced an LTP.
\cite{Blocker_2001} showed that for stars with surface temperatures exceeding 100\,000 K, a VLTP would expose products of nuclear burning, and promptly destroy any remaining hydrogen. 
However, since both models lead to the formation of a giant with a deep convective envelope, further mixing can occur and result in hydrogen deficiency and an enrichment of s-process elements.
Both VLTP and LTP models have been used to explain hydrogen-deficient stars including some central stars of planetary nebulae, PG1159 stars, O(He) stars \citep{miller_06b}, the [WC]-type Wolf-Rayet stars \citep{demarco02}, and R Coronae Borealis (R\,CrB) stars \citep{clayton_1996}. 

\subsection{``Born-again'' stars}

Stars undergoing a Late Thermal Pulse (LTP) are not uncommon: more than 10\% of AGB stars may experience this process \citep{Hans_2003}.

\cite{Iben_1983b} suggest that if the thermal pulse occurs during the Planetary Nebula (PN) phase, the star would almost never linger in the giant phase, but would be observed in the post-AGB phase. 
If the thermal pulse occurs in the white dwarf phase, the star could remain in the giant stage for a period.
During this time, it could be observed and identified as an R\,CrB star, such as FG Sge, V4334 Sgr (Sakurai's Object) and V605 Aql.
FG Sge is a classic demonstration of LTP theory.
\cite{Iben_1983b}’s born-again calculation shows that an LTP model can reach the same luminosity as FG Sge.
Over an interval of 120 years, FG Sge transformed from being a hot post-AGB star into a very luminous cool supergiant. 
During this period, its surface abundance changed from hydrogen-rich to hydrogen-deficient \citep{jeffery06}. 
This rapid reverse evolution and change in surface abundance perfectly aligns with the predictions of the LTP.

The most striking example of an LTP is the central star of the Stingray Nebula \citep{Reindl_2017}. Although it is not hydrogen-deficient, it remains the only LTP object to date that has been observed undergoing both blueward evolution (heating and contraction) and redward evolution (cooling and expansion). Between 1988 and 2002 its surface heated from approximately 38 000 K to 60 000 K and then, beginning in 2002, cooled and expanded, reaching an effective temperature of about 50 000 K and a surface gravity that dropped from log g = 6.0 to 5.5 by 2015.
 
Typical examples of a VLTP are Sakurai's Object and V605 Aql.
\cite{duerberk_1996} presented evidence that Sakurai's Object cooled rapidly and uniformly over time, exhibiting hydrogen deficiency, carbon enrichment, and high luminosity. 
These characteristics indicate that Sakurai's Object underwent a very Late Thermal Pulse. Subsequent observations have shown that the surface hydrogen abundance  declined from approximately 0.007 in May 1996 to about 0.0007 in October 1996\citep{asplund97b}. 
\cite{miller_2007} computed a VLTP model to explain the rapid evolution of Sakurai's Object, achieving an excellent match with the observed cooling rate.

V605 Aql demonstrates the most extreme surface heating rate known, the surface temperature rising from $T_{\rm eff}\approx5\,000$K in 1921 to $\approx95\,000$K in 2006. 
V605 Aql exhibits surface abundances similar to those found in Wolf-Rayet [WC] central stars of planetary nebulae, with about 55\% helium and 40\% carbon \citep{Clayton_2006}. 
Due to the large temperature range and the rapid rate of change, averaging 5\,000 K\,yr$^{-1}$ in the later stages, this is considered a very typical example of a VLTP. 
Due to the striking similarities between Sakurai's Object in 1996 and V605 Aql in 1919, V605 Aql is considered a precursor to the future evolution of Sakurai's Object.

\subsection{DY Cen}

At various epochs, the spectrum of DY\,Cen has resembled a cool hydrogen-deficient star, an extreme helium star having above average surface hydrogen abundance \citep{jeffery93a} and, recently, a late [WC]-type PN central star \citep{jeffery20a}.   
It showed R\,CrB-type variations until 1934, but not subsequently \citep{hoffleit30,schaefer16}. 
Its evolution is rapid; its steady decline in visual brightness over the last century corresponds to an increase of surface temperature  \citep{demarco02,schaefer16}. Combined with spectroscopic measurements of surface temperature covering 1987 - 2015, DY\,Cen has contracted at constant luminosity, its surface heating at a rate of about 300 K yr$^{-1}$ in the last 60 years \citep{pandey14,schaefer16,jeffery20a}.
This rate is a factor 2 higher than expected for regular R CrB and extreme helium stars formed from double white dwarf mergers ($10 - 150$ K\,yr$^{-1}$) \citep{saio_jeffery2002}. 
This contraction is associated with rotational spin-up approaching 50\% of the critical rate \citep{jeffery20a}.
The 1987 spectrum of DY\,Cen showed the surface abundance of the s-process element strontium to be 4 dex above solar \citep{jeffery20a}. 
The latter offers an additional test for evolutionary models of potential LTP stars 
\citep{jeffery06,asplund00}. 
 
\subsection{Objectives}

DY Cen is a rapidly evolving star with reliable surface abundance and long-term observational data, and it shares many similarities with V605 Aql. Both stars are thought to have evolved from the R CrB domain, following similar evolutionary paths and sharing surface characteristics. However, DY Cen also displays a unique surface hydrogen abundance, and its comparison with classical VLTP objects plays a crucial role in the study of born-again stars.

Although several studies have modeled born-again stars, there is still a lack of detailed discussion regarding the conditions that lead to different born-again scenarios, especially VLTP models, and the corresponding range of parameter space. In this work, we use the "Modules for Experiments in Stellar Astrophysics"({\sc mesa}) code to establish a grid of born-again star models under various parameters. We aim to investigate the models' sensitivity to key factors such as the overshoot parameter, stellar mass, and H envelope mass, thereby expanding our understanding of born-again star models.

In particular, we aim to compare these models with the observed characteristics of DY\,Cen to establish whether it could have been produced in a Late or a Very Late Thermal Pulse.

\subsection{Outline}

Section\,\ref{s: methods} of this paper describes the computational tools and physics adopted to build stellar evolution models. 

Section\,\ref{s:models} discusses various model sequences.
Section\,\ref{s:vltp_model} details the process of constructing a standard VLTP model. 
 Section\,\ref{s: class} presents and discusses four models with the same core mass but leading to three distinct  evolutionary outcomes. 
Section\,\ref{s: residual mass} showcases a sequence of models where the hydrogen envelope mass is the variable parameter.
Section\,\ref{core_mass} explores the impact of varying core mass values on surface abundances.
Section\,\ref{overshooting} presents the grid with convective overshooting and illustrates the effects of overshooting on the evolutionary state and surface elemental abundances.

Section \ref{s:dycen} compares our models with observations of DY\,Cen.
 Section\,\ref{s:heating} compares the evolutionary history of DY\,Cen with standard post-AGB and born-again models in terms of heating rates.
 Section\,\ref{s:logg} contrasts the log g-$T_{\rm eff}$ evolutionary tracks of DY\,Cen and VLTP models.

Section \ref{s:conc} summarizes the key findings and broader implications of the study.

\begin{figure}
    \centering
    \includegraphics[width=1\linewidth]{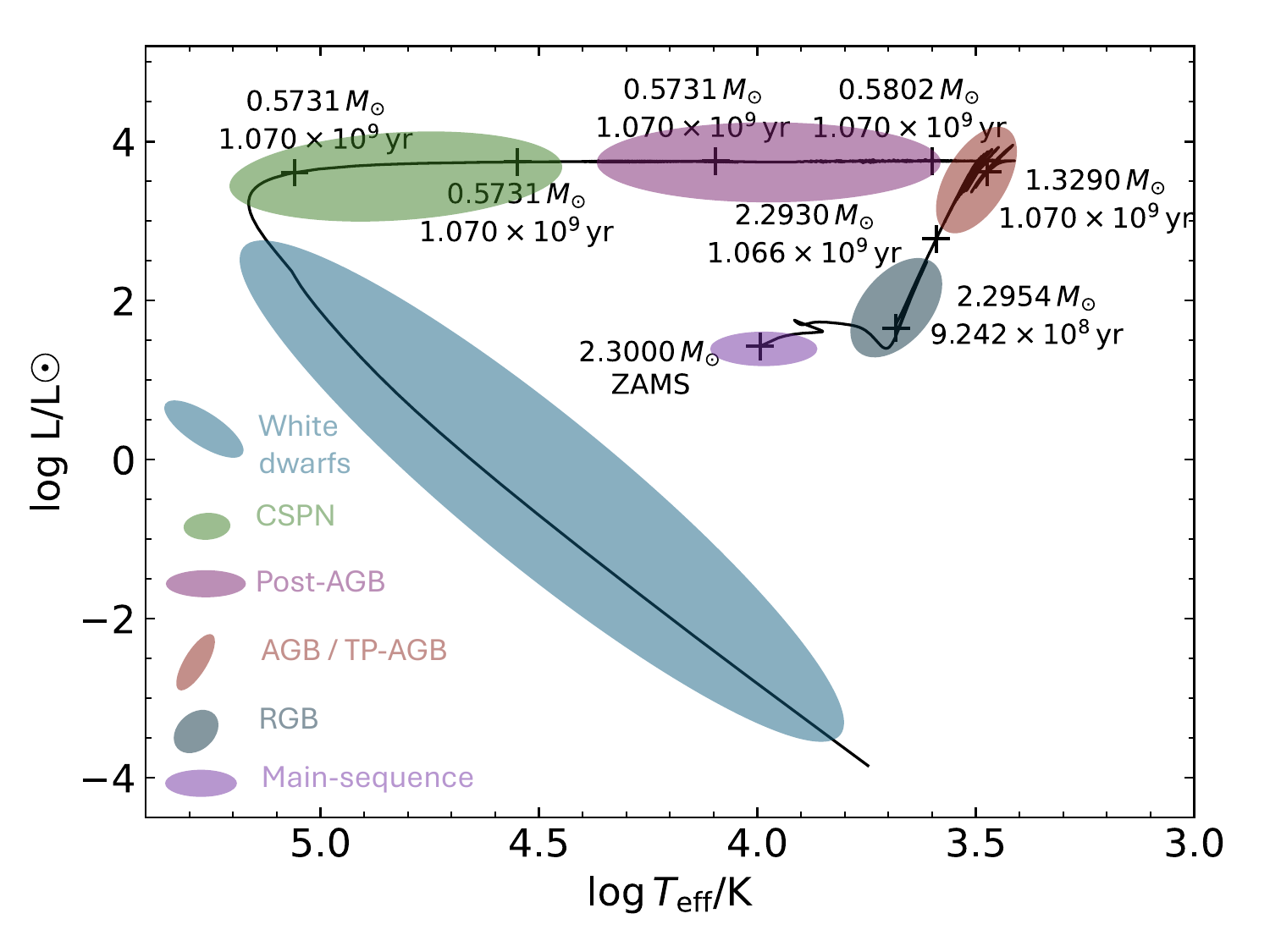}
    \includegraphics[width=1\linewidth]{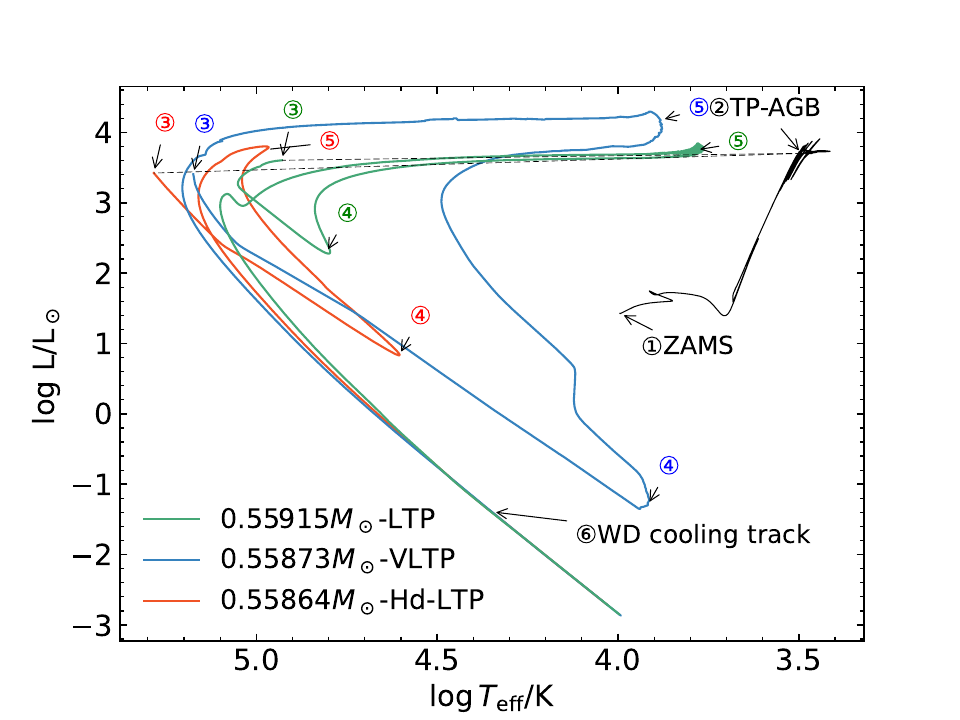}    
    \caption{Top: Classical evolution of 2.3\Msun star ($Z=0.02$) from ZAMS to WD. The main stages of evolution are identified by coloured ellipses, and sample models are labelled with total mass and age from the ZAMS. 
    Bottom: Three different evolutionary tracks calculated for models with the same core mass ($M_{\rm c}=0.55864\Msun$)} but different envelope masses and hence different total masses as shown. Numbers \textcircled{1} to \textcircled{6} identify the main phases of VLTP evolution as in Fig.\,\ref{f:vltp}, distinguished by color. The black numbers represent the stages that are common to all tracks. 
    \label{f:introduce}
\end{figure}

\begin{figure}
\centering
\includegraphics[width=1\linewidth]{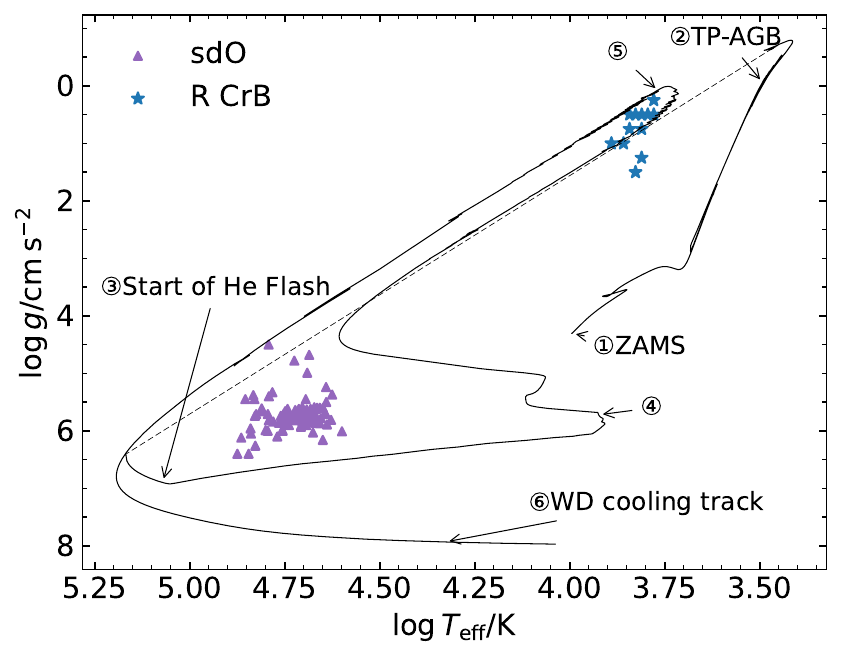}
\caption{ Teff - log g diagram showing the theoretical evolution track for a star that experiences a Very Late Thermal Pulse (VLTP). The rapid mass-loss process is represented by a dashed line.
Numbers \textcircled{1} to \textcircled{6} identify the main phases of VLTP evolution. 
The figure also shows the positions of R CrB stars (in blue, from \cite{simon_2011}) and sdO stars (in purple, from \cite{geier_2022}).
}
\label{f:vltp}
\end{figure}

\section{METHODS} \label{s: methods}

\begin{table*}
\caption{The upper part of the table shows the mass of each component of the standard model (Sect.\,3.1) at four representative stages of evolution. For comparison, the lower part of the table shows the total and hydrogen masses for the same AGB model allowed to evolve to the white dwarf stage without additional mass loss. }
\label{Tab:2}
\begin{center}
\begin{tabular}{ccccc}
\hline\hline
Stage & Total Mass/M$_{\odot}$ & \texttt{H\_mass}/M$_{\odot}$ & \texttt{He\_mass}/M$_{\odot}$ & \texttt{CO\_mass}/M$_{\odot}$ \\
\hline
\multicolumn{4}{l}{Standard model with enhanced mass loss}\\
ZAMS & $2.30000$ & $2.30000$ & $0.00000$ & $0.00000$ \\
TP-AGB & $1.63569$ & $1.07706$ & $0.02944$ & $0.52919$ \\
Blue Point 1 & $0.55875$ & $0.00016$ & $0.02944$ & $0.52915$ \\
Blue Point 2 & $0.55875$ & $0.00000$ & $0.03022$ & $0.52853$ \\
\hline
\multicolumn{4}{l}{Normal AGB and post-AGB evolution with Bl\"ocker wind}\\
ZAMS & $2.30000$ & $2.30000$ & $0.00000$ & $0.00000$ \\
AGB Tip & 1.32901 & 0.75956 & 0.02306 & 0.54639 \\ 
post-AGB cool & 0.58017 & 0.00780 & 0.02473 & 0.54764 \\
post-AGB hot & 0.57315 & 0.00034 & 0.02504 & 0.54777 \\
\hline
\end{tabular}
\end{center}
\end{table*}

We use the stellar evolution code {\sc mesa}-r22.11.1 \citep{Paxton_2011, Paxton_2013, Paxton_2015, Paxton_2018, Paxton_2019} to model the evolution of LTP and VLTP stars. 
Since our born-again star models are initialized at the thermal-pulsing asymptotic giant branch (TP-AGB) stage, we begin by evolving each model from the zero-age main sequence (ZAMS) through to the TP-AGB. In all phases, convective overshooting is included using the  scheme with $f=0.016$ and $f_0 = 0.008$ \citep{herwig00}.
The ratio of mixing length to local pressure scale height is set to $\alpha = l/H_{\rm p} = 1.9179$ \citep{Paxton_2011}. 
Unless otherwise stated, all models in Section\,\ref{s:models} do not include the effects of convective overshooting.
The calculation of opacity uses the default MESA tables as described by \citep{Paxton_2011}  .
At high temperatures ($\log T \geq 3.75$), OPAL Type 2 tables with the  \cite{asplund09} mixture (including C/O enhancement) are selected by setting {\tt kap\_file\_prefix='a09'} and {\tt kap\_CO\_prefix='a09\_co'}.
These are valid for hydrogen-deficient mixtures. 
At low-temperatures ($\log T < 3.75$), the built-in \cite{Ferguson_2005} tables are scaled to the \cite{asplund09} mixture, specified by {\tt kap\_lowT\_prefix='lowT\_fa05\_a09p'.} 

Our post-LTP/VLTP models do not reach surface temperatures $\log T_{\rm eff} < 3.75$ and therefore should not be significantly affected by the low-$T$ Ferguson tables which are only valid for hydrogen-rich mixtures.
The outer boundary condition is chosen to be an Eddington gray photosphere. 
Nuclear reactions are treated with the ‘agb.net’ network, which includes 17 nuclides: \textsuperscript{1}H, \textsuperscript{2}H, \textsuperscript{3}He, \textsuperscript{4}He, \textsuperscript{7}Li, \textsuperscript{7}Be, \textsuperscript{8}B, \textsuperscript{12}C, \textsuperscript{13}C, \textsuperscript{13}N, \textsuperscript{14}N, \textsuperscript{15}N, \textsuperscript{16}O, \textsuperscript{17}O, \textsuperscript{18}O, \textsuperscript{19}F, and \textsuperscript{22}Ne.

All red giants experience substantial mass loss through a stellar wind.
In our AGB models this is achieved through the use of a Bl\"ocker-type formula \citep{Bloecker_1995}.
Fig.\, \ref{f:introduce} (top panel) shows the classical evolution track of a 2.3\Msun\ star with solar composition ($Z=0.02$), and Section \ref{s:vltp_model} will explain this model in detail.
However, to explore a larger range of post-AGB models than allowed by canonical mass-loss formulae,  we used the  "relax\_mass" function in MESA.
This function enforces a rapid stellar wind, artificially driving the loss of the envelope.
This approach allows us to identify the optimal timing for envelope removal -- typically just before the onset of a thermal pulse -- and to control the resulting core mass. 
After numerous attempts, we found that the most consistently successful results were obtained if the outer layers were stripped from the model 
 
The success rate obtained from repeated trials \textbf{indicated} that the optimal moment for envelope removal to eventually trigger a born-again event occurs during a thermal pulse cycle precisely at the point where the helium luminosity begins to rise sharply. This occurs around $\log L_{\rm He}/\Lsun \sim 2.3$, after which the luminosity increases rapidly.

Since the enhanced mass-loss process is carried out whilst the model is on the AGB, the contraction of the residual envelope and relaxation of the underlying layers will bring the model into equilibrium before an LTP or VLTP occurs, so that the model transitions into a pre-white dwarf configuration without helium ignition. 

We note that our approach does not represent a physically continuous evolutionary process. 
Rather, it is designed to produce artificial post-AGB models and hence explore a wider range of possible AGB core and envelope masses than allowed by standard assumptions about AGB mass-loss and evolution. 
In particular we want to systematically identify configurations that are most likely to lead to LTP and VLTP events.
There may be some justification for considering non-canonical mass-loss. 
Other mass-loss mechanisms include  common-envelope ejection or stripping induced by a binary or planetary companion and a super-wind phase leading to the ejection of a planetary nebula.  Although there have been no confirmed binary detections amongst LTP and VLTP candidates, a low-mass or white dwarf companion in a wide binary would be difficult to exclude. 

Note that the envelope has already been enriched with helium and other elements as a consequence of convective dredge-up on both giant branches.
For reference, the total mass (\texttt{Total\_mass}) represents the sum of the envelope mass (\texttt{H\_mass}), the helium-shell mass (\texttt{He\_mass}), and the carbon-oxygen core mass (\texttt{CO\_mass}).
According to \cite{Stancliffe_2005}, approximately $10^{-4}{\rm M}_{\odot}$ hydrogen should remain if an LTP is to occur. 
We will explore the correlation between envelope mass, core mass and the type of late thermal pulse in Section 3.4.

The MESA inlists used to obtain the results presented in this study will be made available elsewhere\footnote{\url{https://zenodo.org/records/15721350}}. Additionally, we have shared the programs developed for creating and processing the model grids, which feature support for a graphical user interface (GUI) mode, making interactions more efficient and intuitive\footnote{\url{https://github.com/584146519/MESAgrid}}.



\begin{figure*}[!htbp]
    \centering
    \includegraphics[width=1\linewidth,trim={0 3cm 0 4cm},clip]{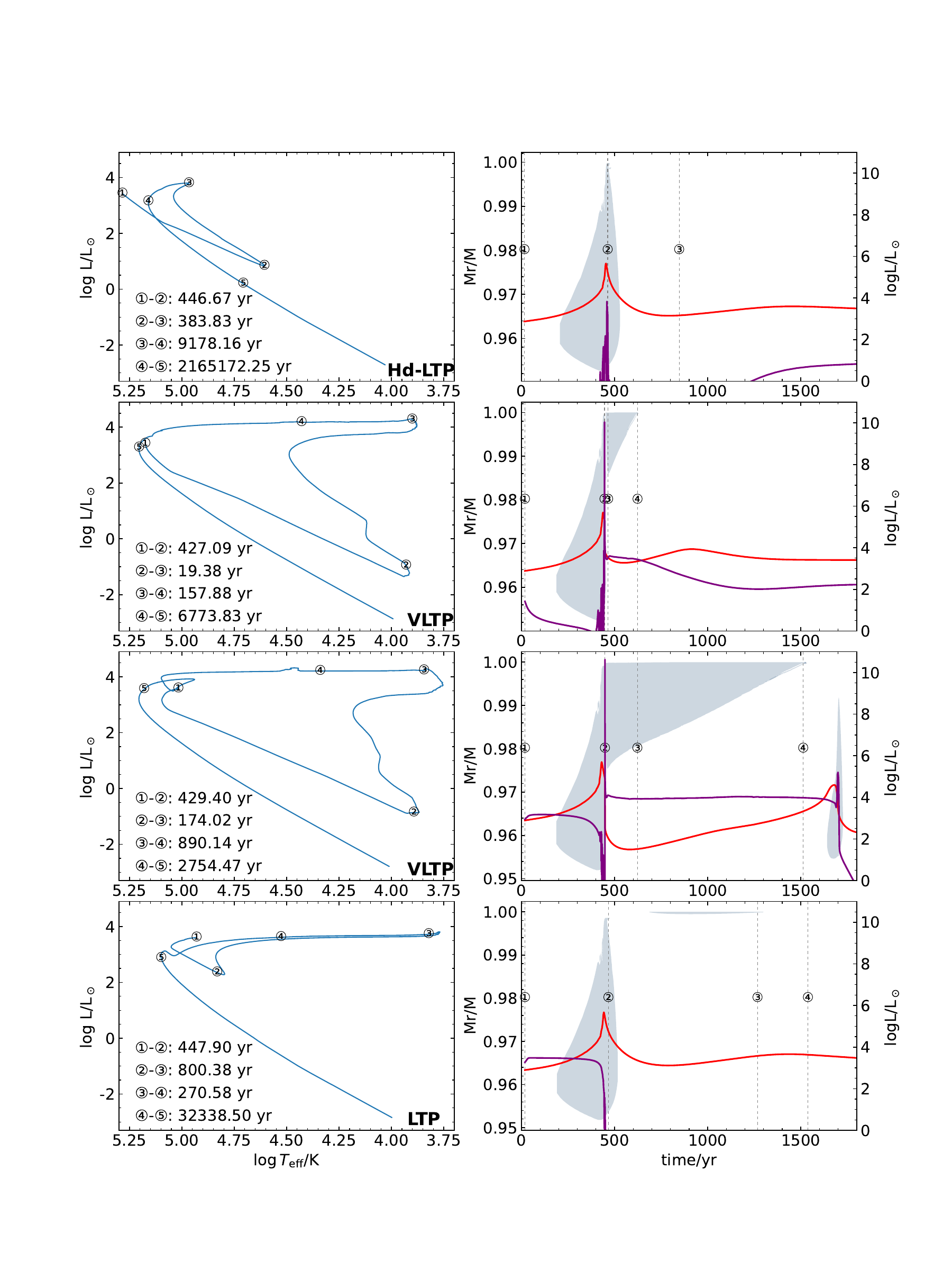}
    \caption{Stellar evolution following a VLTP, i.e. a He-shell flash while on the white dwarf cooling track, for four models with masses  $M = 0.55864, 0.55875, 0.55901$, and $0.55915 \Msun$. All have core mass $M_{\rm c}=0.55864\Msun$. Left: tracks in the $L-T_{\rm eff}$ plane from the end of mass loss (1) to the final white dwarf stage (5). 
    Other phases marked are (2) first red limit and maximum extent of flash-driven convection, (3) maximum radius and (4) $T_{\rm eff}=30\,000$\,K. 
    Right: the evolution of convective regions in terms of mass (shaded) and the luminosities for the H (violet) and He (red) shells for the same models, both as a function of time after the last thermal pulse.}
    \label{f:convHR}
\end{figure*}


\section{Models} \label{s:models}
\subsection{The Standard Model}\label{s:vltp_model}
To ensure that a star undergoes its final thermal pulse during the white dwarf phase, the first step is to model a TP-AGB star and then find the most suitable time to remove most of its envelope just before a thermal pulse occurs. 

We take the example of a $2.3M_{\odot}$ progenitor star with a metallicity of $Z=0.02$ and a helium abundance of $Y=0.28$. 
 
The progenitor star evolves from the pre-main sequence phase, passes through the main sequence and red giant branch (RGB) phases, and undergoes a total of 9 thermal pulses in the thermally pulsing asymptotic giant branch (TP-AGB) phase.\

Evolution after application of the ``relax\_mass'' function marks the beginning of the Very Late Thermal Pulse process we are attempting to model.
For the VLTP model discussed in this section, $\texttt{Total\_mass} = M = 0.55875\Msun$ (the mass after mass loss), $\texttt{H\_mass} = 0.000145 \Msun$ (mass of H envelope), and $\texttt{core\_mass} \equiv \texttt{CO\_mass} + \texttt{He\_mass} = M_{\rm c} = 0.55864\Msun$ (the mass of star without hydrogen envelope).

Table \ref{Tab:2} presents the masses of each component at various evolutionary stages for both our enhanced mass-loss model and the model with a steady, lower-intensity Bl\"ocker wind (the latter using a “Blocker\_scaling\_factor = 0.5").
TP-AGB refers to the last moments of normal evolution before the addition of rapid mass loss. 
The Blue Points (1, 2)  represent the moments when the hottest point was passed before and after the born-again process.

Figure \ref{f:vltp} shows the $\log g - \log T_{\rm eff}$ diagram with the whole evolutionary track of the model from zero-age main sequence (ZAMS) contraction to the final white dwarf cooling track after undergoing a Very Late Thermal Pulse.

The track of the 2.3 M$_\odot$ model starts at the ZAMS.
The first half depicts evolution through the main sequence \textcircled{1}, the first giant branch, helium core burning, the early asymptotic giant branch (E-AGB), and finally, the thermally pulsing asymptotic giant branch (TP-AGB) \textcircled{2}.
Rapid mass loss is imposed during the TP-AGB phase so that the model contracts rapidly to become a white dwarf.

The second half of the track shows the initial white dwarf contraction. 
Soon after the star commences its initial descent on the white dwarf cooling track, a change in slope occurs at the point marked “start of He flash” \textcircled{3}. 
This corresponds to the point where helium shell ignition occurs, and the white dwarf starts to expand. 
Subsequently, the model expands, hydrogen shell ignition occurs \textcircled{4}, and flash-driven convection reaches the surface and enriches surface carbon. 
Further expansion leads to an increase in luminosity and decrease in surface temperature, bringing the model back to the yellow supergiant part of the $\log L-\log T_{\rm eff}$ diagram \textcircled{5}.
Following this, the star begins a new phase of contraction and heating, potentially illuminating a planetary nebula, and will eventually evolve into a carbon-oxygen white dwarf \textcircled{6}.

If we modify the mass loss, thereby altering the mass of the H envelope, we obtain the other two types of Late Thermal Pulse, as illustrated in Figure \ref{f:introduce}.

\subsection{Classification of the Late Thermal Pulse}\label{s: class}

As already introduced, the Late Thermal Pulse can be classified using the location of the thermal pulse which divides pulse into VLTP or LTP.


In our model sequences with different hydrogen‐envelope masses, we identify a case that is clearly distinct from the classical VLTP. In this scenario, the hydrogen envelope is much more depleted than in a standard VLTP, so that a full hydrogen flash cannot develop during the thermal‐pulse. The result is a comparatively smaller loop on the HR diagram than in the VLTP case. We designate this behavior as a hydrogen‐deficient late thermal pulse (Hd-LTP). 
These Hd-LTP models come with a warning. The bottom layers of the H-envelope in AGB stars are very tightly bound. Common-envelope or planetary engulfment events, or other envelope instabilities would be highly unlikely to remove these layers. It is not known what mechanism might remove them in practice. 
However, such artificial models allow us to explore the limits of what might be observed under extreme and hitherto unimagined circumstances.

The evolution tracks for the three classifications are shown in Fig.\, \ref{f:introduce} (lower panel). These are a Hd-LTP model with \texttt{total\_mass} $M = 0.55864\Msun$, a standard VLTP model with $M= 0.55875\Msun$, and a LTP model with $M= 0.55915\Msun$.
Before the final thermal pulse occurred, the models in this subsection all had the same core mass of $M_{\rm c} = 0.55864\Msun$.
In Section \ref{s: residual mass}, the Hd-LTP, VLTP, and LTP are linked into a sequence by treating the hydrogen mass as a continuous variable. 

To understand the entire process, we examined four models to explore the convection zone and luminosity history, corresponding to total mass $M = 0.55864, 0.55873, 0.55901$, and $0.55915 \Msun$.
All have core mass $M_{\rm c}=0.55864\Msun$.
Fig.\,\ref{f:convHR} shows the evolution tracks, extent of convection zones, and evolution of the H envelope and He shell luminosities for each of these models. 

The evolution of the Hd-LTP model can be understood by observing the size and extent of the track loops. 
In the  $M=0.55864$ M$_{\odot}$ model (A), stellar winds have blown away most of the hydrogen before the core cools and contracts to ignite the helium shell.
Therefore there is not enough hydrogen left on the surface to ignite and expand. 
As a result, the model shows a small loop in the Hertzsprung-Russell (HR) diagram, which should becomes larger as the amount of hydrogen increases. 
Fig.\, \ref{f:convHR}(A) shows the $M=0.55864$ M$_{\odot}$ model and that the hydrogen mass is too small to support a high-luminosity ignition.

Figure \ref{f:convHR}(B) shows that the $M=0.55873 \Msun$ model is a VLTP model. As the helium shell luminosity increases, surface convection region also gradually expands, then, hydrogen near the surface ignites, leading to a hydrogen flash. 

It is this hydrogen flash that causes the star to  expand rapidly and, hence, to become brighter. At the same time, a convective region forms on the surface, which connects with the convective region from helium burning, altering the surface element abundance.

When the model evolves back to the hot end, if the surface hydrogen burns to produce enough helium, there will be a secondary brief thermal pulse causing a small loop in the HR diagram, as shown in Fig.\, \ref{f:convHR}(C) for the 0.55901 M$_{\odot}$ model, which is also VLTP model.
This small loop is located in the region where the PG 1159 stars are found \citep{werner_herwig2006}.

Figure \ref{f:convHR}(D) shows that the $M=0.55915\Msun$ model is an LTP model. The surface temperature is lower at the start of the cooling phase, and hydrogen is not ignited throughout the process. Although a thin layer of the convective region forms on the surface during the later stages, there is no contact with the convection zone produced by helium burning, indicating that the LTP does not alter the surface element abundance.

\begin{figure}
    \centering
    \includegraphics[width=1\linewidth]{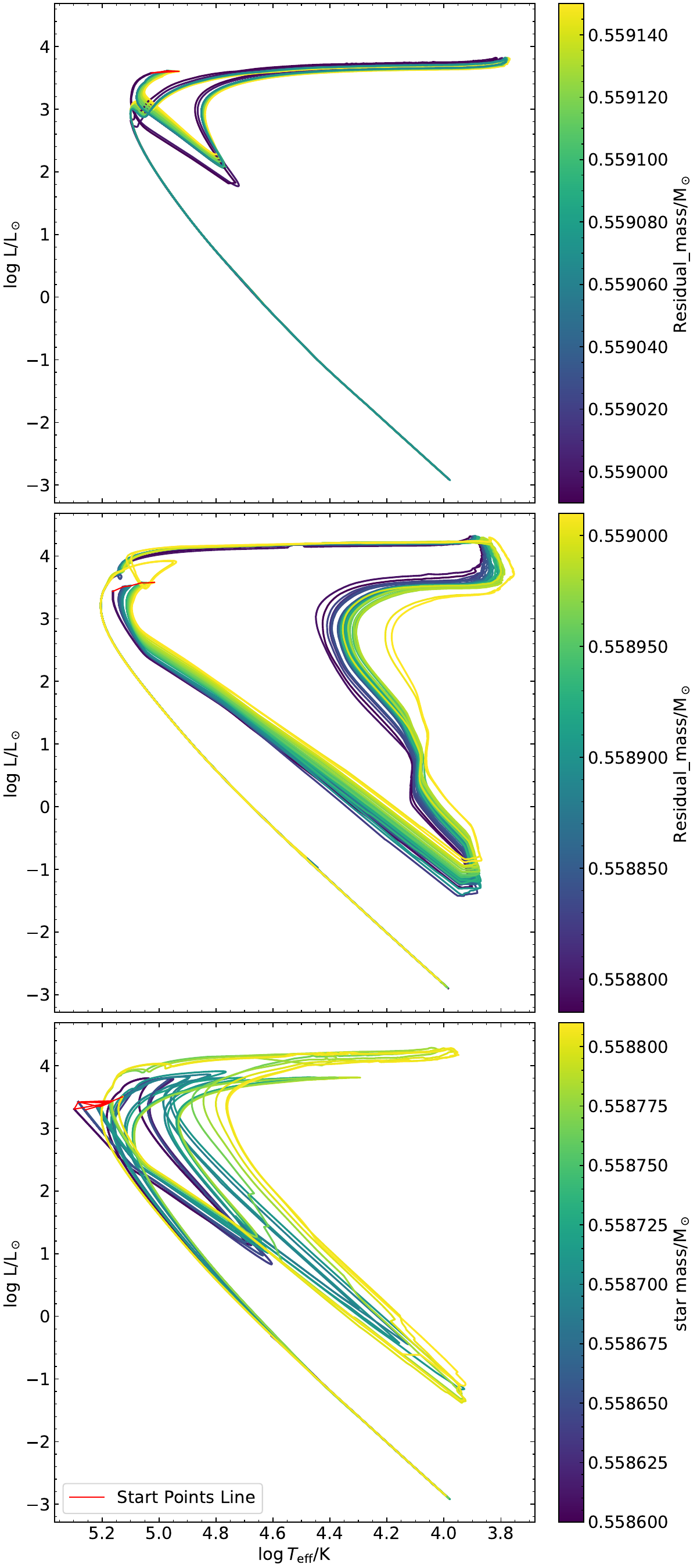}
    \caption{Evolutionary tracks for models having the same initial core mass ($M_{\rm c}=0.55864\Msun$) and with different total masses ($M$) after mass-loss. The red line connects the start positions for the tracks. The tracks are colour-coded by total star mass, as shown in the colour scale on the right.
    Bottom: Hd-LTP models ranging from $M = 0.558605 - 0.558800 \Msun$ from the end of the first post-AGB phase to the white dwarf stage. Middle: VLTP models for $M = 0.558730 - 0.559010 \Msun$.
    Top: LTP models for $M= 0.558980 - 0.559150 \Msun$.
    }
    \label{f:all_shallow}
\end{figure}

\begin{figure}
    \centering
    \includegraphics[width=1\linewidth]{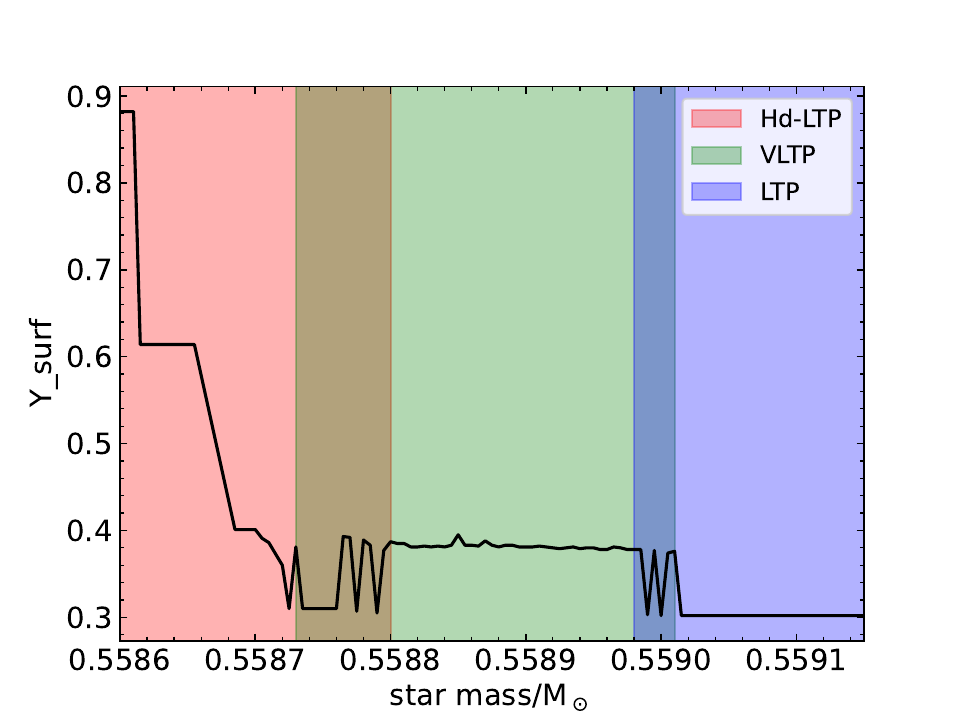}
    \caption{The helium abundance after convective mixing from all ~0.559 \Msun models, with varying hydrogen envelope masses.}
    \label{f:HE_plot}
\end{figure}

\subsection{Evolution as a function of envelope mass}\label{s: residual mass}

 Whilst the intensity of a thermal pulse depends on physical conditions in the inter-shell, the question of whether any
convection driven by the helium shell flash can penetrate to the surface depends on the hydrogen envelope mass lying on top of the inter-shell.    
\citet{lawlor21} analyzed the effect of varying the residual hydrogen mass on LTP models; we extend this analysis to encompass the hydrogen envelope mass range in VLTP models. 
These models do not take into account the effects of convective overshooting.

This section focuses on models created from the 2.3 $M_{\odot}$ progenitor star. The total masses range from $M = 0.558605$ to  $0.559150\Msun$, in intervals of $0.000005\Msun$, where $M = 0.558605 \Msun \equiv M_{\rm c}$ represents the completely stripped core. 
The character of the evolution tracks changes as a function of hydrogen envelope mass (Fig. \ref{f:all_shallow}). The tracks have been divided into three types: Hd-LTP, VLTP, and LTP. Models with larger total mass, corresponding to models with more massive hydrogen envelopes, show changes in the shape of the VLTP loop. Specifically, towards higher mass, the minimum loop luminosity increases, the minimum  loop temperature decreases, and the loop becomes “fatter” overall.
The evolutionary state also changes: what can be considered a Hd-LTP transforms into a true VLTP. The minimum luminosity before expansion can drop as low as $\log L/{\rm L_{\odot}} = -1.4$ .

Figure \ref{f:HE_plot} shows helium surface abundance as a function of total mass during its final post-LTP contraction. 
The three cases are Hd-LTP, VLTP, and LTP.
A Hd-LTP  occurs when the total mass is less than about 0.55872 $M_{\odot}$ and is due to excessive stripping of the surface that exposes the He core. 
The surface helium abundance reflects the composition profile at the hydrogen-helium interface prior to surface stripping.
VLTP corresponds to masses between approximately 0.55875 $M_{\odot}$ and 0.55905 $M_{\odot}$, when a convective hydrogen envelope meets a convective helium shell, dredging elements to the surface. In our models, the surface helium abundance is about 38\%.
The LTP occurs when the mass exceeds about 0.55905 $M_{\odot}$, and it represents an unmixed envelope equivalent to that of the hydrogen-rich envelope after second dredge-up, with a surface helium abundance of about 30\%.

The boundary between VLTP and LTP models is indistinct and sensitive to numerical noise;  a small change in mass or mass-zoning affects the final outcome, particularly within the region where their parameter spaces overlap. 

\begin{table*}
\begin{center}
\caption{Surface abundances as mass fraction (per cent) following a VLTP for models with different core masses. }
\label{tab:massmodels}
\begin{tabular}{ccccccc}
\hline \hline
Initial Mass & Model Mass & H1 (\%) & He4 (\%) & C12 (\%) & N14 (\%) & O16 (\%) \\
\hline
2.30\Msun & 0.55886\Msun & 0.95 & 37.59 & 35.49 & 1.21 & 17.56 \\
2.70\Msun & 0.58770\Msun & 0.26 & 43.56 & 31.88 & 2.17 & 13.96 \\
2.90\Msun & 0.60770\Msun & 0.42 & 44.90 & 31.64 & 2.17 & 12.23 \\
3.10\Msun & 0.63340\Msun & 0.01 & 46.60 & 32.25 & 2.26 & 10.44 \\
3.25\Msun & 0.65870\Msun & 0.88 & 49.17 & 29.10 & 2.70 & 8.10 \\
3.40\Msun & 0.68250\Msun & 0.59 & 51.10 & 30.06 & 2.83 & 5.86 \\
3.60\Msun & 0.72145\Msun & 1.00 & 79.21 & 10.03 & 3.55 & 0.42 \\[2mm]
DY Cen & --- & 2.6(5) & 94(1) & 2.0(9) & 0.13(04) & 0.60(06) \\
\hline
        \end{tabular}
    \end{center}
\end{table*}

\begin{figure}
    \centering
    \includegraphics[width=1\linewidth]{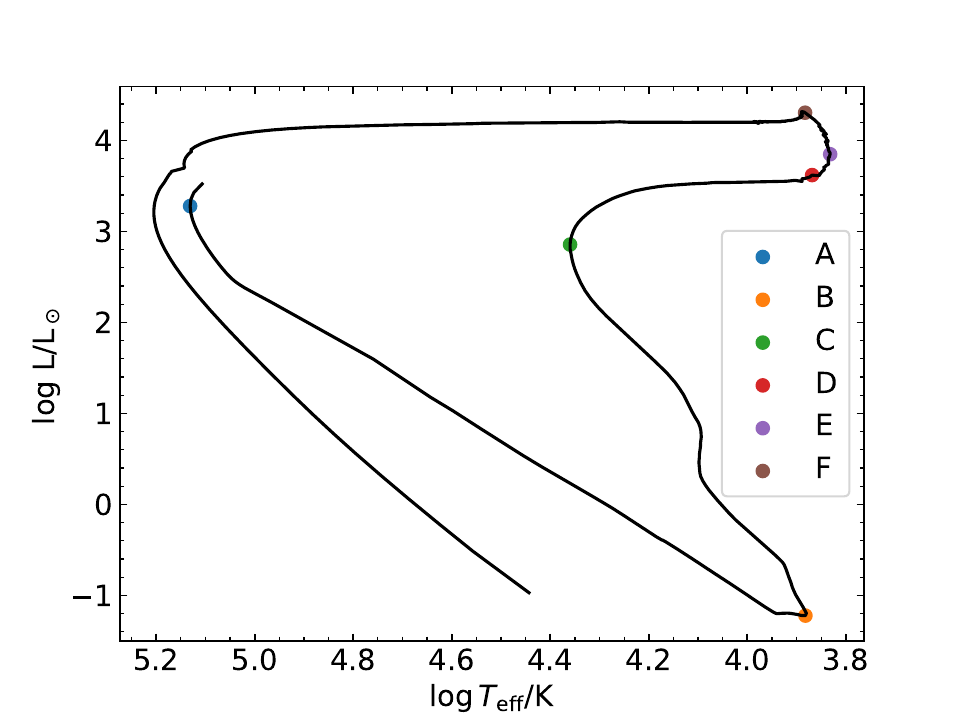}
        \includegraphics[width=1\linewidth]{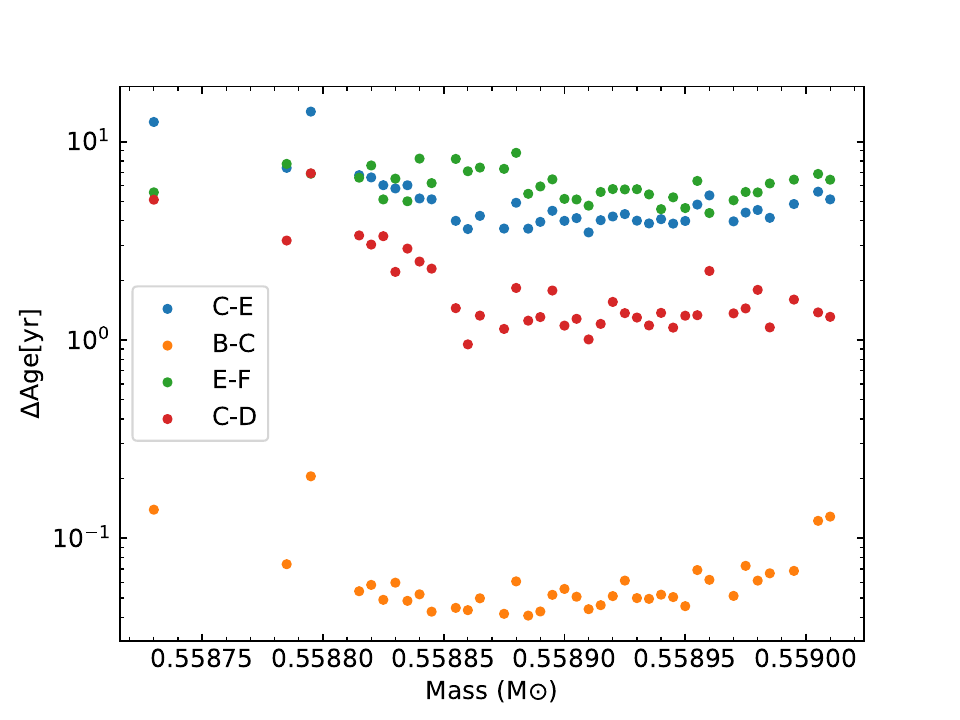}
    \includegraphics[width=1\linewidth]{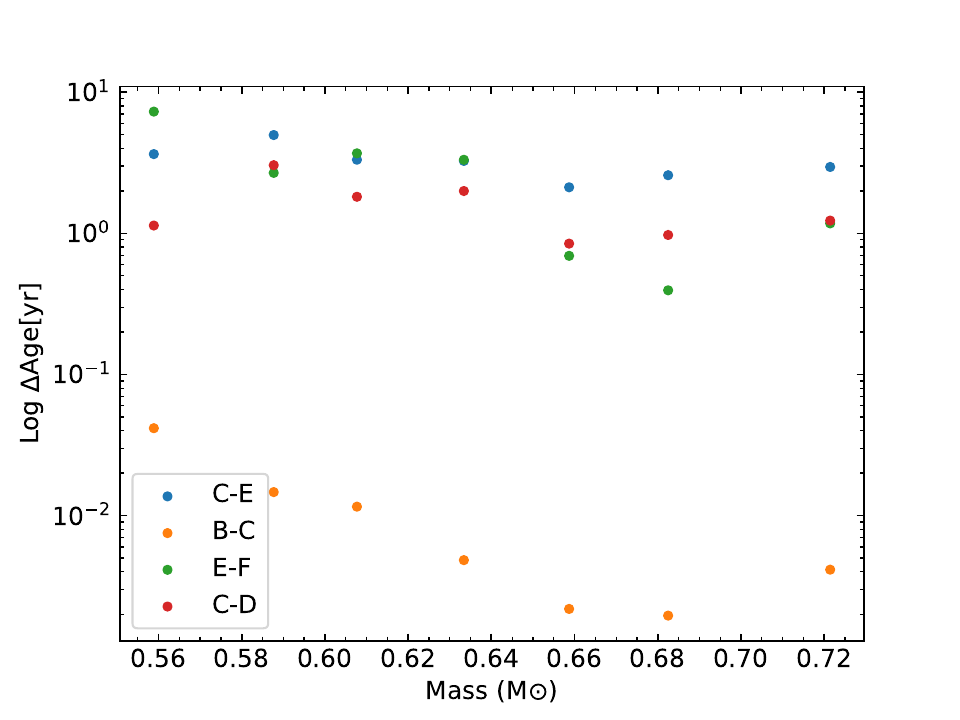}

    \caption{Timescales between critical points on stellar  evolution tracks following a VLTP. 
    Top: The $M = 0.55880 \Msun$ VLTP evolution track is marked to illustrate locations of critical points as defined in the text.
    Middle:  timescales of all VLTP models in the $M_{\rm c}=0.55864 \Msun$ sequence.
    Bottom:  timescales of VLTP models in the varying core mass sequence.}
    
    \label{f:timescale}
\end{figure}

\subsection{Timescales for the Late Thermal Pulse} \label{Timescale}

Figure \ref{f:timescale} shows the elapsed times between critical points on our VLTP evolution tracks. The positions of critical points are illustrated schematically in the top panel and are defined generally as: 
A: maximum effective temperature before the TP,
B: minimum luminosity during the TP,
C: maximum effective temperature during the TP,
D: a point preceding and hotter than E by 500 K, 
E: minimum effective temperature following the TP, and
F: minimum surface gravity following the TP. 
The interval C-D, rather than C-E, better indicates the rate of expansion after the initial pulse, since expansion slows towards the cool limit.  

The lower panels show critical intervals for the sequences of models with the same core mass ($M_{\rm c}=0.55864\Msun$: middle) and increasing envelope mass and with increasing core mass (bottom). 
The initial expansion immediately following shell helium ignition (B-C) lasts $\approx 4$\,d for $M_{\rm c}=0.55864\Msun$ sequence, falling to less than 1\,d at higher core masses. 
For stars like Sakurai's Object,  this would correspond to the pre-discovery phase. 
Expansion from blue-to-red at roughly constant luminosity (C-D) takes between 1 and 5 years, with the shorter times corresponding to larger envelope or core mass. 
For Sakurai's Object this would correspond to the post-discovery phase prior to its obscuration by ejected dust. 

\textbf{The expansion timescales C-D are in good agreement with observations of Sakurai's Object \citep{arkhipova99}.
The timescales C-E are in good agreement with some precious VLTP calculations \citep[{e.g.}][]{miller_2007,lawlor2001}. 
However, other VLTP models require an adjustment of the convective eﬀiciency in order to reproduce such a short timescale \citep[{e.g.}][]{hajduk05,herwig11}.
Our own models indicate that such an adjustment may not be necessary.
}
\subsection{Effect of core mass} \label{core_mass}

We evolved $Z=0.02$ models with initial masses of 2.3, 2.7, 2.9, 3.1, 3.25, and 3.4  $M_{\odot}$ from the PMS stage through the TP-AGB phase, during which they experienced 9, 9, 7, 6, 5, and 3 thermal pulses, respectively, before departing the AGB and eventually becoming born-again models with final masses $M=  0.5587, 0.5877, 0.6077, 0.6334, 0.6586$ and $0.6825\Msun$.
These have core masses $M_{\rm c} =  0.55864, 0.58759, 0.60759, 0.63335, 0.65861$ and $0.68244\Msun$, respectively.

We also evolved a 3.6\,\Msun\ sequence, yielding two models of $M=0.72145\Msun$ and 0.73140\,\Msun, with core masses $M_{\rm c}=  0.72111$  and 0.73128\,\Msun, which underwent one and three thermal pulses (TPs), respectively. 
All models were computed both with and without overshooting, but we excluded models which suffer only one thermal pulse on the TP-AGB without overshoot; their structure leads to anomalous behaviour compared with models which have at least 3 TPs.

\begin{figure}
    \centering
    \includegraphics[width=1\linewidth]{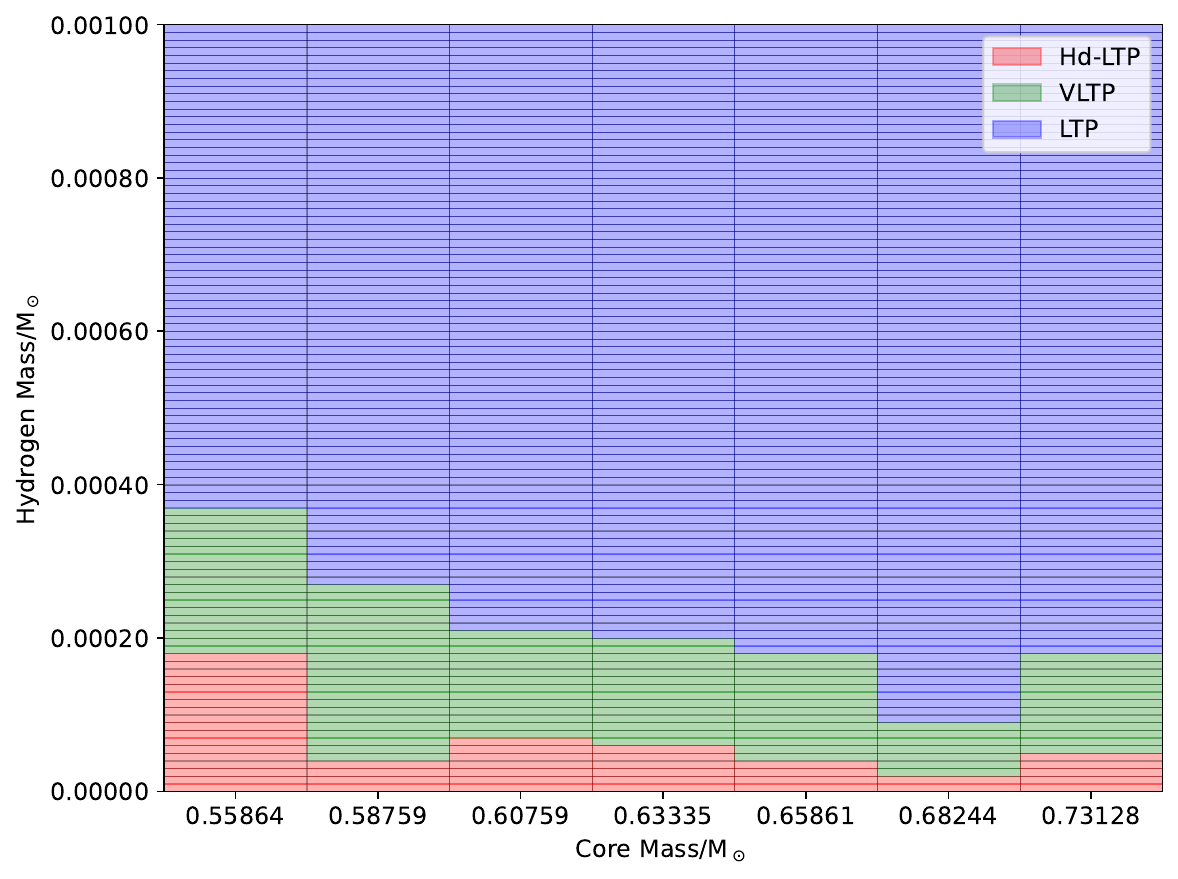}
    \caption{Born-again types as a function of CO+He} core mass and H envelope mass without convective overshooting. 
    \label{f:allmodel}
\end{figure}

\begin{figure}
    \centering
    \includegraphics[width=1\linewidth]{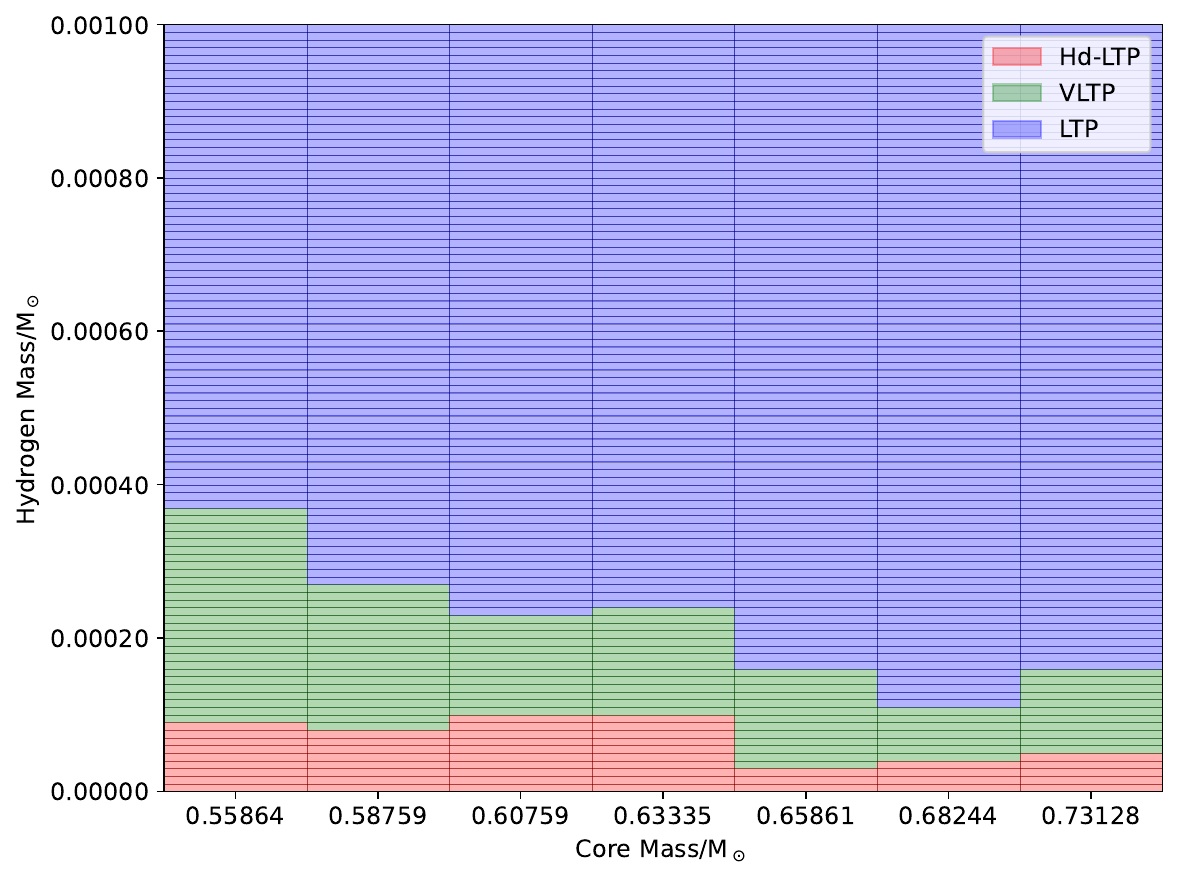}
    \caption{As Fig.\,\ref{f:allmodel} with convective overshooting.}
    \label{f:allmodel-overshoot}
\end{figure}

Figure \ref{f:allmodel} shows the range of different born-again types obtained by varying the hydrogen envelope mass for models with different core masses, without convective overshooting.
Figure \ref{f:allmodel-overshoot} shows the same with convective overshooting included ($f=0.008$). 

\cite{miller_2024} pointed out that more massive remnants tend to possess less massive envelopes when a late thermal pulse occurs. A similar trend can be roughly observed in our model sequence, providing the momdel experiences at least three thermal pulses on the AGB, as shown in Figs.\,\ref{f:allmodel} and \ref{f:allmodel-overshoot}.

Table \ref{tab:massmodels} shows the surface abundance produced by VLTP models with different core masses and without convective overshooting.
It shows that, with increasing core mass, the surface helium (He) abundance increases significantly, while carbon (C) and oxygen (O) abundances gradually decrease. Nitrogen (N) increases relatively slowly, and hydrogen (H) shows no obvious change.

\begin{figure}
    \centering
    \includegraphics[width=1\linewidth]{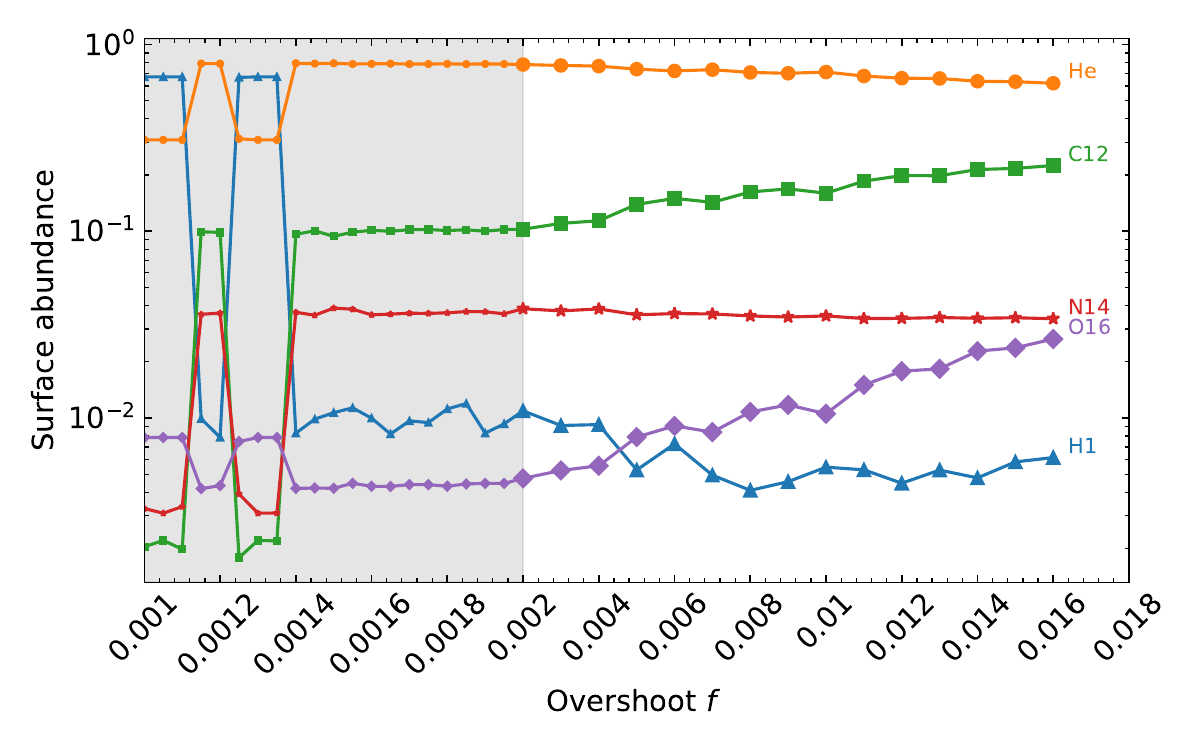}
    \caption{The surface abundances of H, He, C, N and O (mass fraction) for the $M=0.72145\Msun$ model, with convective overshooting parameter $f$ ranging from 0.001 to 0.016.The X-axis of the gray background section uses scales of different proportions to display the data of the more detailed model with $f$ ranging from 0.001 to 0.002}
    \label{f:overshoot}
\end{figure}

\subsection{Effect of overshooting} \label{overshooting}

\citet{Herwig_2000} and \citet{Bloecker_2000} have discussed the significant impact of convective overshooting on surface element abundance during the third dredge-up of AGB stars.
In MESA, the parameter $f_0$ defines how far back into the convection zone to start overshooting, while $f$ specifies the distance the overshooting extends beyond the convection zone edge, both measured in pressure scale heights ($H_p$).

 In order to investigate the influence of convective overshooting on surface abundance and evolution, we built a grid for a $M=0.72145 \Msun$ model, where we systematically varied the convective overshooting parameters. Specifically, we set the parameter $f_0$ to 0.001 and adjusted $f$ to explore different degrees of overshooting.

\citet{Fei} argue that setting $f=0.008$ may be  appropriate for low-mass main-sequence models.   However, the same value cannot be assumed  for all  stages of stellar evolution. 

Using $f=0.008$ as a reference, we created a series of convective overshooting models to explore the range $f =  0 - 0.016$. The step size from 0 to 0.002 is 0.00005, and the step size from 0.002 to 0.016 is 0.001.

Figure \ref{f:overshoot} shows the relationship between the convective overshooting parameter $f$ and surface abundance for the $M=0.72145\Msun$ model. When $f \sim  0.00005-0.0012$, the model behaves mostly as a Late Thermal Pulse (LTP). 
For $f > 0.0012$, the model behaves as a Very Late Thermal Pulse (VLTP)
The transition occurs because convection brings hydrogen into the convective zone, triggering a hydrogen ignition that leads to a VLTP. 
Subsequently, as overshooting increases, the surface abundance of helium decreases while that of carbon and oxygen increases. 
This is due to the greater entrainment of CO core material into the helium convection zone, which is then carried to the surface by the convective zone of the subsequent hydrogen flash, thereby altering the abundances of helium, carbon and oxygen.
In this case (Fig.\,\ref{f:overshoot}), the LTP/VLTP transition is bistable in the range $f \sim  0.00005-0.0012$, with solutions oscillating between LTP and VLTP behaviour. Whilst the bistability suggests numerical noise, increasing the temporal, spatial and model-grid resolution have all, so far, failed to resolve the bistability.


\begin{figure}
    \centering
    \includegraphics[width=1\linewidth]{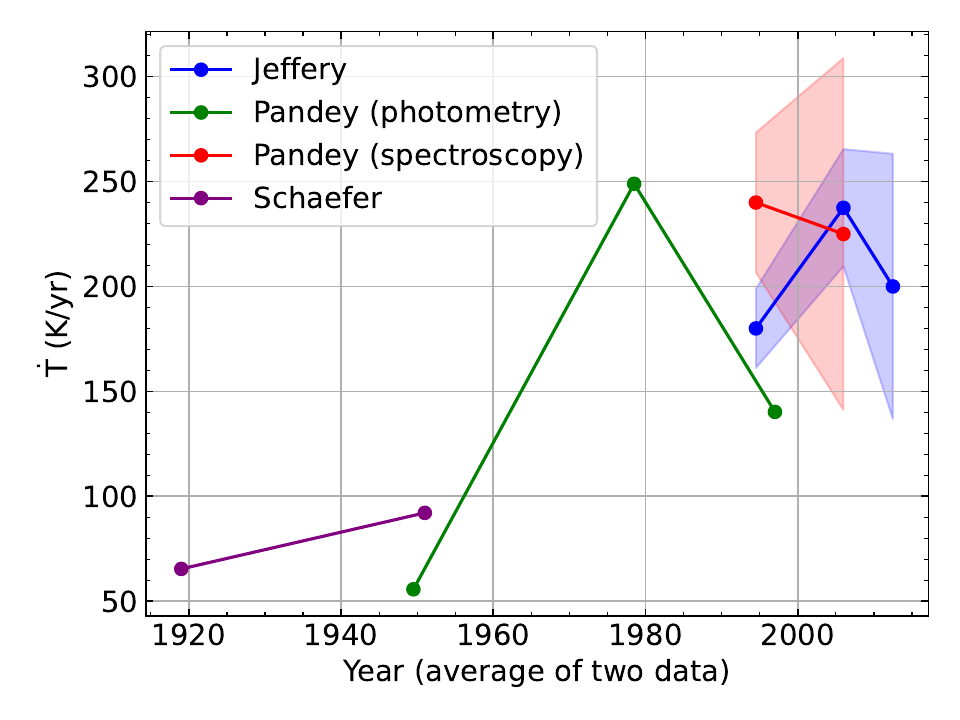}
    \caption{The observed rate of change of surface temperature ($\dot{T}$) for DY\,Cen between 1920 and the present. Values are calculated for the midpoint between two observations using data reported by \citet{schaefer16}, \citet{pandey14} and \citet{jeffery20a}. }
    \label{f:tdot_obs}
\end{figure}

\begin{figure}
    \centering
    \includegraphics[width=1\linewidth]{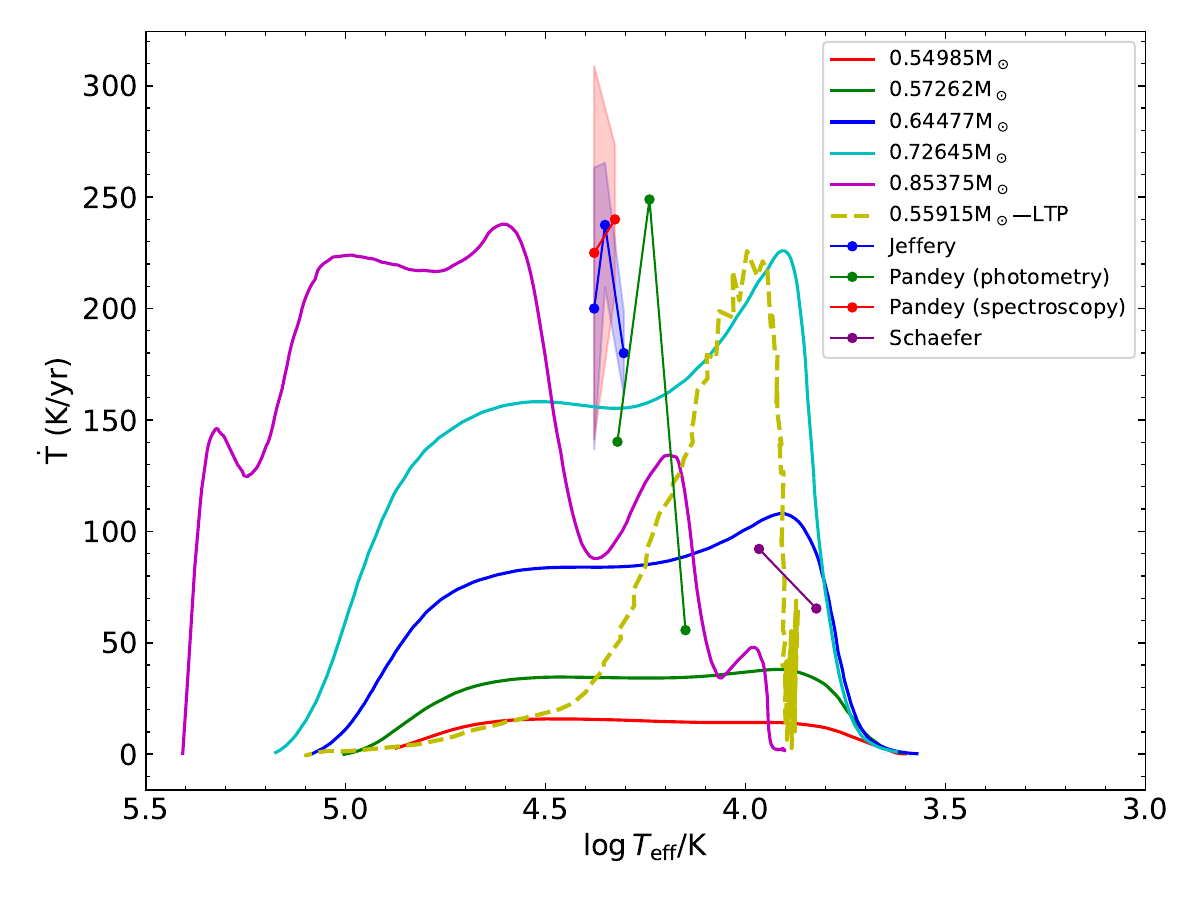}
    \caption{The $T-\dot{T}$ curves for standard stellar models of different masses during the post-AGB phase, with the dashed line representing the curve of the 0.55915 \Msun\ LTP model. 
The DY\,Cen observations from Fig.\,\ref{f:tdot_obs} are included. }
    \label{f: standard Tdot}
\end{figure}

\begin{figure}
    \centering
    \includegraphics[width=1\linewidth]{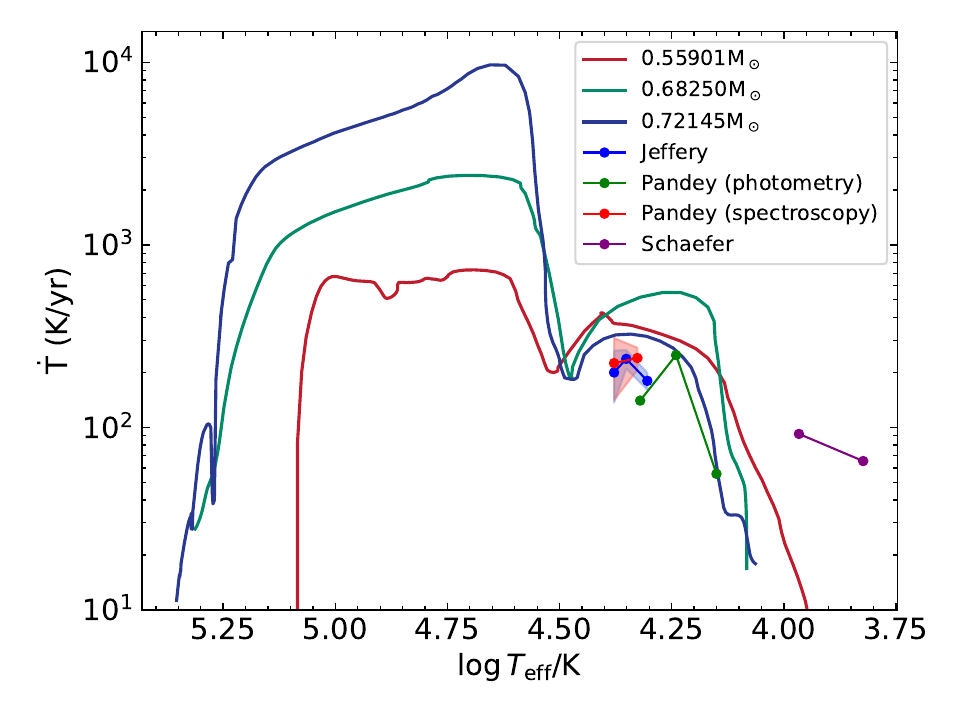}
    \caption{The $T-\dot{T}$ curves for the VLTP (Very Late Thermal Pulse) models of 0.55901 $M_{\odot}$, 0.68250 $M_{\odot}$, and 0.72145 $M_{\odot}$ are shown. In each case, the dashed lines of the same color represent the average  \(\dot{T}\) values for that mass model within the corresponding temperature range.
}
    \label{f:VLTP}
\end{figure}

\section{DY Centauri: case-study of a born-again star} \label{s:dycen}

\subsection{Comparison with observed heating} \label{s:heating}

\citet{Clayton_2006} identified V605\,Aql as a VLTP born-again star. 
Initially identified as a nova-like outburst, 
V605\,Aql brightened rapidly to 12th magnitude in 1921 and 1923, when it had a spectral type R0, similar to some hydrogen-deficient carbon or R\,CrB stars.  
It is now the hydrogen-deficient central star of a planetary nebula,
so \citet{Clayton_2006} deduced that it expanded rapidly following a VLTP, and has now contracted again.    
The surface temperature was around 5000 K in 1921, over 50\,000 K in 1997, and reached $95\,000 {\rm K} \pm 10\,000 {\rm K}$ in 2006 \citep{Clayton_2006}.
On average, the rate of temperature change  $\dot{T}\equiv{\rm d}T_{\rm eff}/{\rm d}t$ from 1921 to 1997 was greater than 592 K\,yr$^{-1}$, and from 1997 to 2006 it accelerated to about 5000 K\,yr$^{-1}$.

This may be compared with DY\,Cen, which was observed to show R\,CrB-like variations with an estimated $T_{\rm eff}\approx 5000$K in 1932 but is now contracting to reach $T_{\rm eff}\approx 24\,000$K in 2016 \citep{jeffery20a}. 
If we calculate the average  $\dot{T}$  between consecutive temperature measurements for DY\,Cen, using data from the same researchers, we can observe the trend of  $\dot{T}$ shown in Fig.\,\ref{f:tdot_obs}. 
DY\,Cen’s surface is heating at an accelerating rate from about 65 K\,yr$^{-1}$ in 1920 to approximately 240 K\,yr$^{-1}$ by 2010. 
Compared to V605\,Aql, DY\,Cen’s evolution is significantly slower. 

DY\,Cen has also been identified as an extreme helium (EHe) star \citep{jeffery93a}, for which the merger of a helium white dwarf with a carbon-oxygen white dwarf (CO+He WD) provides a convincing model \citep{saio_jeffery2002,zhang14}. 
However the observed heating rate for DY\,Cen is  more than a factor of 2 greater than observed for any other EHe star \citep{jeffery01b} or predicted by any CO+He WD merger model \citep{saio_jeffery2002,zhang14}. 
Moreover, DY\,Cen shows a surface hydrogen abundance more than $100\times$ greater (by number) than any other EHe star \citep{jeffery93a,pandey14,jeffery20a}. 
Its surface fluorine and strontium abundances  are enhanced by $\sim2$ and $\sim4$ dex respectively \citep{pandey14,jeffery20a}, reflecting an s-processed intershell on the AGB. 
Its transition to a [WC] type spectrum \citep{jeffery20a} argues for a comparison with the hot RCrB star V348\,Sgr, but this helps  little; neither does the presence of a nebula \citep{demarco02}.  

Consequently it has been appropriate to ask whether there are circumstances under which an LTP or VLTP model could resolve some questions about the origin of DY\,Cen.

We generated several models illustrating the $T-\dot{T}$  evolution during the post-AGB phase (Fig.\,\ref{f: standard Tdot}). These models follow a trajectory typical of classical post-AGB evolution after the star has lost most of its hydrogen envelope during the AGB phase, resulting in a surface hydrogen mass of $\sim 0.02 \Msun$.

The standard post-AGB model is strongly affected by mass. As the mass increases from 0.54985 $M_{\odot}$ to 0.85375 $M_{\odot}$, the maximum value of  \(\dot{T}\) rises from approximately 15 K\,yr$^{-1}$ to around 240 K\,yr$^{-1}$\citep{Paczynski1970}. Additionally, we included the 0.55915 M$_{\odot}$ LTP (Late Thermal Pulse) model for comparison in the plot. The LTP model exhibits a heating rate of about 240 K\,yr$^{-1}$ during the early post-AGB phase, which is nearly an order of magnitude faster than the standard model of the same mass. However, this rate is still far below the 5000 K\,yr$^{-1}$ observed in V605\,Aql.

In the VLTP model, MESA simulations exhibited computational instabilities during the contraction phase, leading to numerically noisy evolution tracks. 
Several attempts to reduce this noise were attempted, including increasing and decreasing the resolution both in space and time, without success.
The tracks shown in Fig.\,\ref{f:VLTP} have been smoothed to show the overall trends.
Fig.\,\ref{f:VLTP} also shows that the heating rate of the VLTP model is  highly sensitive to mass. At $T_{\rm eff}\approx 40\,000$ K,  $\dot{T}$ is often $> 1000$ K\,yr$^{-1}$ and, for the 0.72145 M$_{\odot}$ model,  $\dot{T}$ is closer to 10\,000 K\,yr$^{-1}$. The average  $\dot{T}$ in the temperature range  50\,000 -- 100\, 000 K matches the 5000 K\,yr$^{-1}$ rate observed in V605\,Aql.

\begin{figure}
    \centering
    \includegraphics[width=1\linewidth]{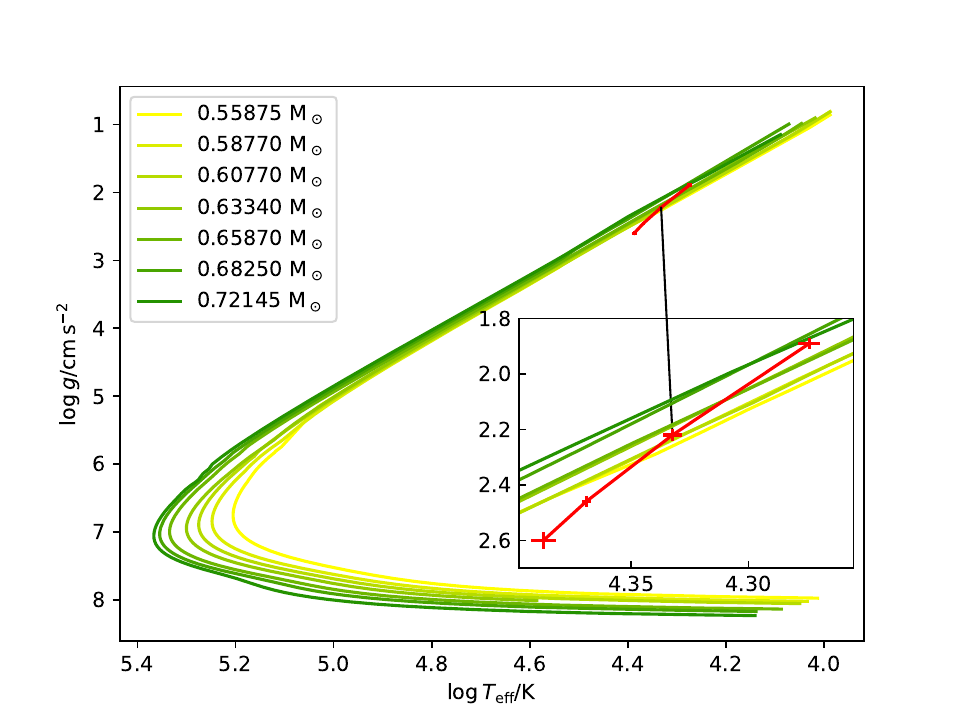}
    \caption{The $\log g - T_{\text{eff}}$ evolutionary tracks for all VLTP models during their final evolution to the white dwarf stage.
    Each track is colour-coded for total mass as shown in the legend. 
    The red curve (with error bars) represents the observations of DY Cen from 1987 - 2015 \citep{jeffery20a}. The inset expands the region around DY\,Cen.  }
    \label{f:logg_teff}
\end{figure}

\subsection{Comparison with observed surface gravity}\label{s:logg}

Figure \ref{f:logg_teff} presents the evolutionary tracks of VLTP models with different masses, from  final contraction to the white dwarf stage, plotted on the same $\log g-T_{\rm eff}$ diagram alongside observations of DY\,Cen \citep{jeffery20a}. Both surface gravity and luminosity are strongly correlated with stellar radius. For a given star, an increase in radius is generally accompanied by a rise in luminosity and always by a decrease in surface gravity, and vice versa. As a result, Fig.\, \ref{f:logg_teff} displays a pattern similar to that of an HR diagram, where the evolutionary tracks of higher-mass stars are located above those of lower-mass stars.
We can use these tracks to constrain the mass predicted by the model to around 0.6 \(M_{\odot}\). The observed track of DY\,Cen appears to be offset by an angle relative to the model tracks. As for the Stingray central star \citep{Reindl_2017}, this may simply be due to systematic errors in the spectroscopic analyses. 

\subsection{Comparison with observed surface composition}

Table\,\ref{tab:massmodels} compares the observed surface composition of DY\,Cen with VLTP models for different core masses. 
DY Cen has a surface dominated by helium at 94\% by mass fraction, with hydrogen at 2.6\%, carbon at 2\% and nitrogen and oxygen at 0.13\% and 0.60\% respectively. 
Such abundances cannot be satisfactorily reproduced by VLTP models, which dredge up carbon- and nitrogen-rich intershell material via a thermal pulse.  Even in a $0.72 {\rm M_{\odot}}$ VLTP model, the surface helium mass fraction peaks at only 79.2\%, falling well short of the observed 94\%.
Consequently, DY\,Cen is too H-rich and C+N+O-poor to match the VLTP models, which all dredge CO-core material to the surface.


\section{ DISCUSSION AND CONCLUSION} \label{s:conc}

Over the past half-century, the Very Late Thermal Pulse (VLTP) and Late Thermal Pulse (LTP) models have provided the leading explanations for the rapid evolution of chemically-peculiar post-Asymptotic Giant Branch (post-AGB) stars. 
In particular, several extremely fast-evolving stars have been confirmed as products of a VLTP, including Sakurai's Object \citep{asplund00} and V605\,Aql \citep{Clayton_2006}. 
Both of these may be related to the R Coronae Borealis (R CrB) variables. 
We have established a comprehensive database of VLTP/LTP models, taking into account various parameters such as core mass, convective overshooting, and envelope mass. 
We have examined key observable properties including surface abundances, heating rates, and evolutionary conditions. 
We have compared the models with the rapidly evolving star DY\,Cen, which also recently passed through the R CrB phase, and have attempted to shed light on its prior evolution.

\subsection{LTP and VLTP models.}

We have systematically examined how LTP and VLTP models behave and how this depends primarily on the hydrogen (H) envelope mass remaining after the star leaves the AGB for the first time: 

1. When the H envelope mass is approximately between 0 and \( 5 \times 10^{-5} M_\odot \), the star undergoes a Hydrogen‐deficient Late Thermal Pulse (Hd-LTP). In this case, the star burns the last of its helium (He) and remaining H during the early white dwarf stage, creating a very small loop in its evolutionary track. 
These Hd-LTP models are highly exploratory; there is no known mechanism which could remove the most tightly bound layers of hydrogen and hence create the necessary initial conditions.

2. When the H envelope mass is between \( 5 \times 10^{-5} M_\odot \) and \( 1 \times 10^{-3} M_\odot \), the star experiences a convective zone near the surface, driven by He burning in the early white dwarf phase. This triggers a hydrogen flash, which results in a final dredge-up of surface material. This process produces the largest loop on the Hertzsprung-Russell (HR) diagram, representing a classic VLTP.

3. When the H envelope mass exceeds \( 1 \times 10^{-3} M_\odot \), the He-shell flash driven convection zone does not reach the  H-rich surface layers and hence there are no changes in surface composition.
Following He-shell ignition, the He-rich layers expand to such an extent that the H-shell is extinguished. 
In this case, the star returns to the AGB phase through stable He burning, and this Late Thermal Pulse (LTP) appears as a relatively small, flat loop in the HR diagram.\\

4. As already established, the heating rate is mass-sensitive \citep{Reindl_2017,lawlor21,miller_06b}. 
This is systematically illustrated using $T-\dot{T}$ diagrams for VLTP models of different masses. In their contraction phases, LTP and VLTP models evolve faster than standard post-AGB models by approximately one and two orders of magnitude, respectively. 

Convective overshooting (Section \ref{overshooting}) affects the models in the following ways:
As the convective overshooting parameter increases in VLTP models, the abundance of helium (He) gradually decreases, while the abundances of carbon (C) and oxygen (O) increase. The overall effect of increasing convective overshooting on surface abundances exhibits a trend opposite to that observed with increasing stellar mass.

\subsection{DY\,Cen}

Considering that DY\,Cen is a rapidly evolving hydrogen-deficient star with both detailed surface abundance measurements and nearly a century of observations, it holds particular significance. 
Crucially, in the early 20th century, DY\,Cen had a similar temperature to V605\,Aql, and both were reported as R\,CrB stars at the time. 
Given that V605\,Aql has already been confirmed as a VLTP star, DY\,Cen is a rare and key object for validating evolution to the R CrB phase and subsequent contraction to become an  H-deficient hot white dwarf. 

Since DY\,Cen shows a surface hydrogen abundance $100\times$ higher and a contraction rate at least 2$\times$ higher than observed in stars thought to arise from a CO+He WD merger, the question arises whether a VLTP model could be helpful. 
We have therefore compared DY Cen observations with our LTP and VLTP models. 
From the perspective of surface abundances, the standard LTP model predicts a relatively hydrogen-rich surface typical for a post-AGB star.  With substantial convective overshoot \citep[e.g. $f=0.064$ ][]{Blocker_2001},  final dredge-up can occur and enrich the surface with helium, carbon and heavier elements comparable with the VLTP models in Table\,\ref{Tab:2}. 
Neither LTP case is compatable with the observations of DY\,Cen. 
Similarly, the extremely high helium abundance observed in DY\,Cen (94\%) is unlikely to be realized with a VLTP model; even the 0.72\Msun VLTP model has only 79.2\% helium. 
Based on the observed track in $T_{\rm eff} - \log g$ space, a VLTP model for DY\,Cen would have a mass around 0.6\,\Msun. 
However, the corresponding VLTP contraction rate ($\approx 1\,000$ K\,yr$^{-1}$) is much higher than that observed over the last 120 years.
From these perspectives, the DY\,Cen observations do not satisfactorily match any of our LTP or VLTP models, particularly with regard to the elemental abundances. 

In summary, both LTP and VLTP models, including the extreme Hd-VLTP models, predict hydrogen and carbon abundances which are much higher than observed and must be ruled out, for now. The merging CO+He WD model leads to enhanced carbon and nitrogen abundances at the level observed in DY\,Cen. High fluorine and strontium abundances can be explained by material from the AGB intershell of the CO white dwarf precursor and from the ring of fire associated with the merge itself.  
The challenge is to find a combination of parameters within the CO+He model which retains sufficient hydrogen from the progenitor white dwarfs, as well as providing for rapid contraction after the RCrB phase has completed.


\section*{Acknowledgments}
We thank the referee for the helpful suggestions and comments that improved the manuscript. This work is supported by the grants 12473028 and 12073006 from the National Natural Science Foundation of China. Armagh Observatory and Planetarium is funded by the Northern Ireland Department for Communities. 

\vspace{5mm}


\bibliography{jerrybib}{}

\begin{thebibliography}{}
\expandafter\ifx\csname natexlab\endcsname\relax\def\natexlab#1{#1}\fi
\providecommand{\url}[1]{\href{#1}{#1}}
\providecommand{\dodoi}[1]{doi:~\href{http://doi.org/#1}{\nolinkurl{#1}}}
\providecommand{\doeprint}[1]{\href{http://ascl.net/#1}{\nolinkurl{http://ascl.net/#1}}}
\providecommand{\doarXiv}[1]{\href{https://arxiv.org/abs/#1}{\nolinkurl{https://arxiv.org/abs/#1}}}

\bibitem[{V.~P. {Arkhipova} {et~al.}(1999){Arkhipova}, {Noskova}, {Esipov}, \& {Sokol}}]{arkhipova99}
{Arkhipova}, V.~P., {Noskova}, R.~I., {Esipov}, V.~F., \& {Sokol}, G.~V. 1999, \bibinfo{title}{{Sakurai's object (V4334 Sgr) in 1998: The R CrB phase has set in},} Astronomy Letters, 25, 615

\bibitem[{M. {Asplund} {et~al.}(2009){Asplund}, {Grevesse}, {Sauval}, \& {Scott}}]{asplund09}
{Asplund}, M., {Grevesse}, N., {Sauval}, A.~J., \& {Scott}, P. 2009, \bibinfo{title}{{The Chemical Composition of the Sun},} \araa, 47, 481, \dodoi{10.1146/annurev.astro.46.060407.145222}

\bibitem[{M. {Asplund} {et~al.}(1997){Asplund}, {Gustafsson}, {Lambert}, \& {Kameswara Rao}}]{asplund97b}
{Asplund}, M., {Gustafsson}, B., {Lambert}, D.~L., \& {Kameswara Rao}, N. 1997, \bibinfo{title}{{A stellar endgame - the born-again Sakurai's object.},} \aap, 321, L17

\bibitem[{M. {Asplund} {et~al.}(2000){Asplund}, {Gustafsson}, {Lambert}, \& {Rao}}]{asplund00}
{Asplund}, M., {Gustafsson}, B., {Lambert}, D.~L., \& {Rao}, N.~K. 2000, \bibinfo{title}{{The R Coronae Borealis stars - atmospheres and abundances},} \aap, 353, 287

\bibitem[{T. {Bl{\"o}cker}(1995){Bl{\"o}cker}}]{Bloecker_1995}
{Bl{\"o}cker}, T. 1995, \bibinfo{title}{{Stellar evolution of low- and intermediate-mass stars. II. Post-AGB evolution.},} \aap, 299, 755

\bibitem[{T. {Bl{\"o}cker}(2001){Bl{\"o}cker}}]{Blocker_2001}
{Bl{\"o}cker}, T. 2001, \bibinfo{title}{{Evolution on the AGB and beyond: on the formation of H-deficient post-AGB stars},} \apss, 275, 1, \dodoi{10.1023/A:1002777931450}

\bibitem[{T. {Bl{\"o}cker} {et~al.}(2000){Bl{\"o}cker}, {Herwig}, \& {Driebe}}]{Bloecker_2000}
{Bl{\"o}cker}, T., {Herwig}, F., \& {Driebe}, T. 2000, \bibinfo{title}{{AGB evolution with overshoot : hot bottom burning and dredge up},} \memsai, 71, 711, \dodoi{10.48550/arXiv.astro-ph/0002455}

\bibitem[{G.~C. {Clayton}(1996){Clayton}}]{clayton_1996}
{Clayton}, G.~C. 1996, \bibinfo{title}{{The R Coronae Borealis Stars},} \pasp, 108, 225, \dodoi{10.1086/133715}

\bibitem[{G.~C. {Clayton} {et~al.}(2006){Clayton}, {Kerber}, {Pirzkal}, {De Marco}, {Crowther}, \& {Fedrow}}]{Clayton_2006}
{Clayton}, G.~C., {Kerber}, F., {Pirzkal}, N., {et~al.} 2006, \bibinfo{title}{{V605 Aquilae: The Older Twin of Sakurai's Object},} \apjl, 646, L69, \dodoi{10.1086/506593}

\bibitem[{O. {De Marco} {et~al.}(2002){De Marco}, {Clayton}, {Herwig}, {Pollacco}, {Clark}, \& {Kilkenny}}]{demarco02}
{De Marco}, O., {Clayton}, G.~C., {Herwig}, F., {et~al.} 2002, \bibinfo{title}{{What Are the Hot R Coronae Borealis Stars?},} \aj, 123, 3387

\bibitem[{H.~W. {Duerbeck} \& S. {Benetti}(1996){Duerbeck} \& {Benetti}}]{duerberk_1996}
{Duerbeck}, H.~W., \& {Benetti}, S. 1996, \bibinfo{title}{{Sakurai's Object---A Possible Final Helium Flash in a Planetary Nebula Nucleus},} \apjl, 468, L111+

\bibitem[{J.~W. {Ferguson} {et~al.}(2005){Ferguson}, {Alexander}, {Allard}, {Barman}, {Bodnarik}, {Hauschildt}, {Heffner-Wong}, \& {Tamanai}}]{Ferguson_2005}
{Ferguson}, J.~W., {Alexander}, D.~R., {Allard}, F., {et~al.} 2005, \bibinfo{title}{{Low-Temperature Opacities},} \apj, 623, 585, \dodoi{10.1086/428642}

\bibitem[{S. {Geier} {et~al.}(2022){Geier}, {Dorsch}, {Pelisoli}, {Reindl}, {Heber}, \& {Irrgang}}]{geier_2022}
{Geier}, S., {Dorsch}, M., {Pelisoli}, I., {et~al.} 2022, \bibinfo{title}{{Radial velocity variability and the evolution of hot subdwarf stars},} \aap, 661, A113, \dodoi{10.1051/0004-6361/202143022}

\bibitem[{F. {Guo} \& Y. {Li}(2019){Guo} \& {Li}}]{Fei}
{Guo}, F., \& {Li}, Y. 2019, \bibinfo{title}{{Convective Overshooting in Low-mass Stars Using the k-{\ensuremath{\omega}} Model},} \apj, 879, 86, \dodoi{10.3847/1538-4357/ab262f}

\bibitem[{M. {Hajduk} {et~al.}(2005){Hajduk}, {Zijlstra}, {Herwig}, {van Hoof}, {Kerber}, {Kimeswenger}, {Pollacco}, {Evans}, {Lop{\'e}z}, {Bryce}, {Eyres}, \& {Matsuura}}]{hajduk05}
{Hajduk}, M., {Zijlstra}, A.~A., {Herwig}, F., {et~al.} 2005, \bibinfo{title}{{The Real-Time Stellar Evolution of Sakurai's Object},} Science, 308, 231, \dodoi{10.1126/science.1108953}

\bibitem[{F. {Herwig}(2000){Herwig}}]{herwig00}
{Herwig}, F. 2000, \bibinfo{title}{{The evolution of AGB stars with convective overshoot},} \aap, 360, 952

\bibitem[{F. {Herwig} \& T. {Bl{\"o}cker}(2000){Herwig} \& {Bl{\"o}cker}}]{Herwig_2000}
{Herwig}, F., \& {Bl{\"o}cker}, T. 2000, in Liege International Astrophysical Colloquia, Vol.~35, Liege International Astrophysical Colloquia, ed. A.~{Noels}, P.~{Magain}, D.~{Caro}, E.~{Jehin}, G.~{Parmentier}, \& A.~A. {Thoul}, 59, \dodoi{10.48550/arXiv.astro-ph/9909503}

\bibitem[{F. {Herwig} {et~al.}(2011){Herwig}, {Pignatari}, {Woodward}, {Porter}, {Rockefeller}, {Fryer}, {Bennett}, \& {Hirschi}}]{herwig11}
{Herwig}, F., {Pignatari}, M., {Woodward}, P.~R., {et~al.} 2011, \bibinfo{title}{{Convective-reactive Proton-$^{12}$C Combustion in Sakurai's Object (V4334 Sagittarii) and Implications for the Evolution and Yields from the First Generations of Stars},} \apj, 727, 89, \dodoi{10.1088/0004-637X/727/2/89}

\bibitem[{D. {Hoffleit}(1930){Hoffleit}}]{hoffleit30}
{Hoffleit}, D. 1930, \bibinfo{title}{{Variables in Milky Way Field 167},} Harvard College Observatory Bulletin, 874, 13

\bibitem[{I. {Iben}(1984){Iben}}]{Iben_1984b}
{Iben}, Jr., I. 1984, \bibinfo{title}{{On the frequency of planetary nebula nuclei powered by helium burningand on the frequency of white dwarfs with hydrogen-deficient atmospheres.},} \apj, 277, 333, \dodoi{10.1086/161700}

\bibitem[{I. {Iben, Jr.} {et~al.}(1983){Iben, Jr.}, {Kaler}, {Truran}, \& {Renzini}}]{Iben_1983b}
{Iben, Jr.}, I., {Kaler}, J.~B., {Truran}, J.~W., \& {Renzini}, A. 1983, \bibinfo{title}{{On the evolution of those nuclei of planetary nebulae that experiencea final helium shell flash.},} \apj, 264, 605, \dodoi{10.1086/160631}

\bibitem[{I. {Iben, Jr.} \& A.~V. {Tutukov}(1984){Iben, Jr.} \& {Tutukov}}]{iben84}
{Iben, Jr.}, I., \& {Tutukov}, A.~V. 1984, \bibinfo{title}{{Supernovae of type I as end products of the evolution of binaries with components of moderate initial mass (M not greater than about 9 solar masses)},} \apjs, 54, 335

\bibitem[{C.~S. {Jeffery} \& U. {Heber}(1993){Jeffery} \& {Heber}}]{jeffery93a}
{Jeffery}, C.~S., \& {Heber}, U. 1993, \bibinfo{title}{{Spectral analysis of DY Centauri, a hot R Coronae Borealis star with unusually high hydrogen content},} \aap, 270, 167

\bibitem[{C.~S. {Jeffery} {et~al.}(2011){Jeffery}, {Karakas}, \& {Saio}}]{simon_2011}
{Jeffery}, C.~S., {Karakas}, A.~I., \& {Saio}, H. 2011, \bibinfo{title}{{Double white dwarf mergers and elemental surface abundances in extreme helium and R Coronae Borealis stars},} \mnras, 414, 3599, \dodoi{10.1111/j.1365-2966.2011.18667.x}

\bibitem[{C.~S. {Jeffery} {et~al.}(2020){Jeffery}, {Rao}, \& {Lambert}}]{jeffery20a}
{Jeffery}, C.~S., {Rao}, N.~K., \& {Lambert}, D.~L. 2020, \bibinfo{title}{{SALT revisits DY Cen: a rapidly evolving strontium-rich single helium star},} \mnras, 493, 3565, \dodoi{10.1093/mnras/staa406}

\bibitem[{C.~S. {Jeffery} \& D. {Sch\"{o}nberner}(2006){Jeffery} \& {Sch\"{o}nberner}}]{jeffery06}
{Jeffery}, C.~S., \& {Sch\"{o}nberner}, D. 2006, \bibinfo{title}{{Stellar archaeology: the evolving spectrum of FG Sagittae},} \aap, 459, 885

\bibitem[{C.~S. {Jeffery} {et~al.}(2001){Jeffery}, {Woolf}, \& {Pollacco}}]{jeffery01b}
{Jeffery}, C.~S., {Woolf}, V.~M., \& {Pollacco}, D.~L. 2001, \bibinfo{title}{{Time-resolved spectral analysis of the pulsating helium star V652 Her},} \aap, 376, 497

\bibitem[{S. {Kwok}(2024){Kwok}}]{Kwok2024}
{Kwok}, S. 2024, \bibinfo{title}{{Planetary Nebulae Research: Past, Present, and Future},} Galaxies, 12, 39, \dodoi{10.3390/galaxies12040039}

\bibitem[{T.~M. {Lawlor}(2021){Lawlor}}]{lawlor21}
{Lawlor}, T.~M. 2021, \bibinfo{title}{{New models for the rapid evolution of the central star of the Stingray Nebula},} \mnras, 504, 667, \dodoi{10.1093/mnras/stab890}

\bibitem[{T.~M. {Lawlor} \& J. {MacDonald}(2001){Lawlor} \& {MacDonald}}]{lawlor2001}
{Lawlor}, T.~M., \& {MacDonald}, J. 2001, in Astronomical Society of the Pacific Conference Series, Vol. 226, 12th European Workshop on White Dwarfs, ed. J.~L. {Provencal}, H.~L. {Shipman}, J.~{MacDonald}, \& S.~{Goodchild}, 20

\bibitem[{M.~M. {Miller Bertolami}(2016){Miller Bertolami}}]{miller_2016}
{Miller Bertolami}, M.~M. 2016, \bibinfo{title}{{New models for the evolution of post-asymptotic giant branch stars and central stars of planetary nebulae},} \aap, 588, A25, \dodoi{10.1051/0004-6361/201526577}

\bibitem[{M.~M. {Miller Bertolami}(2024){Miller Bertolami}}]{miller_2024}
{Miller Bertolami}, M.~M. 2024, \bibinfo{title}{{Primer on Formation and Evolution of Hydrogen-Deficient Central Stars of Planetary Nebul{\ae} and Related Objects},} Galaxies, 12, 83, \dodoi{10.3390/galaxies12060083}

\bibitem[{M.~M. {Miller Bertolami} \& L.~G. {Althaus}(2006){Miller Bertolami} \& {Althaus}}]{miller_06b}
{Miller Bertolami}, M.~M., \& {Althaus}, L.~G. 2006, \bibinfo{title}{{Full evolutionary models for PG 1159 stars. Implications for the helium-rich O(He) stars},} \aap, 454, 845, \dodoi{10.1051/0004-6361:20054723}

\bibitem[{M.~M. {Miller Bertolami} \& L.~G. {Althaus}(2007){Miller Bertolami} \& {Althaus}}]{miller_2007}
{Miller Bertolami}, M.~M., \& {Althaus}, L.~G. 2007, \bibinfo{title}{{The born-again (very late thermal pulse) scenario revisited: the mass of the remnants and implications for V4334 Sgr},} \mnras, 380, 763, \dodoi{10.1111/j.1365-2966.2007.12115.x}

\bibitem[{B. {Paczy{\'n}ski}(1970){Paczy{\'n}ski}}]{Paczynski1970}
{Paczy{\'n}ski}, B. 1970, \bibinfo{title}{{Evolution of Single Stars. I. Stellar Evolution from Main Sequence to White Dwarf or Carbon Ignition},} \actaa, 20, 47

\bibitem[{B. {Paczy{\'n}ski}(1971){Paczy{\'n}ski}}]{paczynski71}
{Paczy{\'n}ski}, B. 1971, \bibinfo{title}{{Evolution of Single Stars. IV. Helium Stars},} \actaa, 21, 1

\bibitem[{G. {Pandey} {et~al.}(2014){Pandey}, {Rao}, {Jeffery}, \& {Lambert}}]{pandey14}
{Pandey}, G., {Rao}, N.~K., {Jeffery}, C.~S., \& {Lambert}, D.~L. 2014, \bibinfo{title}{{On the Binary Helium Star DY Centauri: Chemical Composition and Evolutionary State},} \apj, 793, 76, \dodoi{10.1088/0004-637X/793/2/76}

\bibitem[{B. {Paxton} {et~al.}(2011){Paxton}, {Bildsten}, {Dotter}, {Herwig}, {Lesaffre}, \& {Timmes}}]{Paxton_2011}
{Paxton}, B., {Bildsten}, L., {Dotter}, A., {et~al.} 2011, \bibinfo{title}{{Modules for Experiments in Stellar Astrophysics (MESA)},} \apjs, 192, 3, \dodoi{10.1088/0067-0049/192/1/3}

\bibitem[{B. {Paxton} {et~al.}(2013){Paxton}, {Cantiello}, {Arras}, {Bildsten}, {Brown}, {Dotter}, {Mankovich}, {Montgomery}, {Stello}, {Timmes}, \& {Townsend}}]{Paxton_2013}
{Paxton}, B., {Cantiello}, M., {Arras}, P., {et~al.} 2013, \bibinfo{title}{{Modules for Experiments in Stellar Astrophysics (MESA): Planets, Oscillations, Rotation, and Massive Stars},} \apjs, 208, 4, \dodoi{10.1088/0067-0049/208/1/4}

\bibitem[{B. {Paxton} {et~al.}(2015){Paxton}, {Marchant}, {Schwab}, {Bauer}, {Bildsten}, {Cantiello}, {Dessart}, {Farmer}, {Hu}, {Langer}, {Townsend}, {Townsley}, \& {Timmes}}]{Paxton_2015}
{Paxton}, B., {Marchant}, P., {Schwab}, J., {et~al.} 2015, \bibinfo{title}{{Modules for Experiments in Stellar Astrophysics (MESA): Binaries, Pulsations, and Explosions},} \apjs, 220, 15, \dodoi{10.1088/0067-0049/220/1/15}

\bibitem[{B. {Paxton} {et~al.}(2018){Paxton}, {Schwab}, {Bauer}, {Bildsten}, {Blinnikov}, {Duffell}, {Farmer}, {Goldberg}, {Marchant}, {Sorokina}, {Thoul}, {Townsend}, \& {Timmes}}]{Paxton_2018}
{Paxton}, B., {Schwab}, J., {Bauer}, E.~B., {et~al.} 2018, \bibinfo{title}{{Modules for Experiments in Stellar Astrophysics (MESA): Convective Boundaries, Element Diffusion, and Massive Star Explosions},} \apjs, 234, 34, \dodoi{10.3847/1538-4365/aaa5a8}

\bibitem[{B. {Paxton} {et~al.}(2019){Paxton}, {Smolec}, {Schwab}, {Gautschy}, {Bildsten}, {Cantiello}, {Dotter}, {Farmer}, {Goldberg}, {Jermyn}, {Kanbur}, {Marchant}, {Thoul}, {Townsend}, {Wolf}, {Zhang}, \& {Timmes}}]{Paxton_2019}
{Paxton}, B., {Smolec}, R., {Schwab}, J., {et~al.} 2019, \bibinfo{title}{{Modules for Experiments in Stellar Astrophysics (MESA): Pulsating Variable Stars, Rotation, Convective Boundaries, and Energy Conservation},} \apjs, 243, 10, \dodoi{10.3847/1538-4365/ab2241}

\bibitem[{N. {Reindl} {et~al.}(2017){Reindl}, {Rauch}, {Miller Bertolami}, {Todt}, \& {Werner}}]{Reindl_2017}
{Reindl}, N., {Rauch}, T., {Miller Bertolami}, M.~M., {Todt}, H., \& {Werner}, K. 2017, \bibinfo{title}{{Breaking news from the HST: the central star of the Stingray Nebula is now returning towards the AGB},} \mnras, 464, L51, \dodoi{10.1093/mnrasl/slw175}

\bibitem[{H. {Saio} \& C.~S. {Jeffery}(2002){Saio} \& {Jeffery}}]{saio_jeffery2002}
{Saio}, H., \& {Jeffery}, C.~S. 2002, \bibinfo{title}{{Merged binary white dwarf evolution: rapidly accreting carbon-oxygen white dwarfs and the progeny of extreme helium stars},} \mnras, 333, 121, \dodoi{10.1046/j.1365-8711.2002.05384.x}

\bibitem[{B.~E. {Schaefer}(2016){Schaefer}}]{schaefer16}
{Schaefer}, B.~E. 2016, \bibinfo{title}{{All known hot RCB stars are fading fast over the last century},} \mnras, 460, 1233, \dodoi{10.1093/mnras/stw1065}

\bibitem[{D. {Sch\"{o}nberner}(1979){Sch\"{o}nberner}}]{schoenberner79}
{Sch\"{o}nberner}, D. 1979, \bibinfo{title}{{Asymptotic giant branch evolution with steady mass loss},} \aap, 79, 108

\bibitem[{R.~J. {Stancliffe}(2005){Stancliffe}}]{Stancliffe_2005}
{Stancliffe}, R.~J. 2005, PhD thesis, University of Cambridge, UK

\bibitem[{H. {van Winckel}(2003){van Winckel}}]{Hans_2003}
{van Winckel}, H. 2003, \bibinfo{title}{{Post-AGB Stars},} \araa, 41, 391, \dodoi{10.1146/annurev.astro.41.071601.170018}

\bibitem[{K. {Werner} \& F. {Herwig}(2006){Werner} \& {Herwig}}]{werner_herwig2006}
{Werner}, K., \& {Herwig}, F. 2006, \bibinfo{title}{{The Elemental Abundances in Bare Planetary Nebula Central Stars and the Shell Burning in AGB Stars},} \pasp, 118, 183, \dodoi{10.1086/500443}

\bibitem[{X. {Zhang} {et~al.}(2014){Zhang}, {Jeffery}, {Chen}, \& {Han}}]{zhang14}
{Zhang}, X., {Jeffery}, C.~S., {Chen}, X., \& {Han}, Z. 2014, \bibinfo{title}{{Post-merger evolution of carbon-oxygen + helium white dwarf binaries and the origin of R Coronae Borealis and extreme helium stars},} \mnras, 445, 660, \dodoi{10.1093/mnras/stu1741}

\end{thebibliography}
\bibliographystyle{aasjournal}
\end{document}